%% file: main.tex
\documentclass[journal]{IEEEtran}
\ifCLASSINFOpdf
\else
\fi
\usepackage{booktabs}
\usepackage{multirow}
\usepackage{graphicx}
\usepackage{subfigure}
\usepackage{amssymb}
\usepackage{mathtools}
\usepackage{float}
\usepackage{color}
\usepackage{url} 
\usepackage{siunitx}
\begin{document}
%
\title{Towards Extreme Image Compression with Latent Feature Guidance and Diffusion Prior}
%
%
%
\author{Zhiyuan Li, Yanhui Zhou, Hao Wei, \textit{Graduate Student Member, IEEE}, Chenyang Ge, Jingwen Jiang
\thanks{This work is supported by the National Natural Science Foundation of China (NSFC62088102, NSFC62376208), China Telecom Group Corporation-Xi’an Jiaotong University Jointly Established Intelligent Cloud Network Science and Education Integration Innovation Research Institute (No.20221279-ZKT03). (\emph{Corresponding author: Chenyang Ge.})}

\thanks{The authors are with the National Key Laboratory of Human-Machine Hybrid Augmented Intelligence, Institute of Artificial Intelligence and Robotics, Xi’an Jiaotong University, Xi’an 710049, China (e-mail: lizhiyuan2839@163.com; zhouyh@mail.xjtu.edu.cn; haowei@stu.xjtu.edu.cn; cyge@mail.xjtu.edu.cn; jiangjingwen@stu.xjtu.edu.cn).}
}

\maketitle

\input{01_abstract}
\input{02_introduction}
\input{03_relatedwork}

\input{04_algorithm}
\input{05_experiments}

\input{06_analysis}

\input{07_conclusion}

\bibliographystyle{unsrt}
\bibliography{reference}
\clearpage
\input{09_biography}

\end{document}

%% file: 01_abstract.tex
\begin{abstract}
Image compression at extremely low bitrates (below 0.1 bits per pixel (bpp)) is a significant challenge due to substantial information loss. In this work, we propose a novel two-stage extreme image compression framework that exploits the powerful generative capability of pre-trained diffusion models to achieve realistic image reconstruction at extremely low bitrates.
In the first stage, we treat the latent representation of images in the diffusion space as guidance, employing a VAE-based compression approach to compress images and initially decode the compressed information into content variables. The second stage leverages pre-trained stable diffusion to reconstruct images under the guidance of content variables. Specifically, we introduce a small control module to inject content information while keeping the stable diffusion model fixed to maintain its generative capability.
Furthermore, we design a space alignment loss to force the content variables to align with the diffusion space and provide the necessary constraints for optimization.
Extensive experiments demonstrate that our method significantly outperforms state-of-the-art approaches in terms of visual performance at extremely low bitrates. The source code and trained models are available at \url{https://github.com/huai-chang/DiffEIC}.

\end{abstract}


%% file: 02_introduction.tex
\section{Introduction}
\begin{figure*}[htbp]\scriptsize
\centering
\makebox[0.196\textwidth]{(a) \textbf{Original}}
\makebox[0.196\textwidth]{(b) \textbf{VVC, 0.0205 bpp}}
\makebox[0.196\textwidth]{(c) \textbf{MS-ILLM, 0.0447 bpp}}
\makebox[0.196\textwidth]{(d) \textbf{Text+Sketch, 0.0281 bpp}}
\makebox[0.196\textwidth]{(e) \textbf{DiffEIC (Ours), 0.0201 bpp}}
\\ \vspace{0.1cm}
\includegraphics[width=0.196\textwidth]{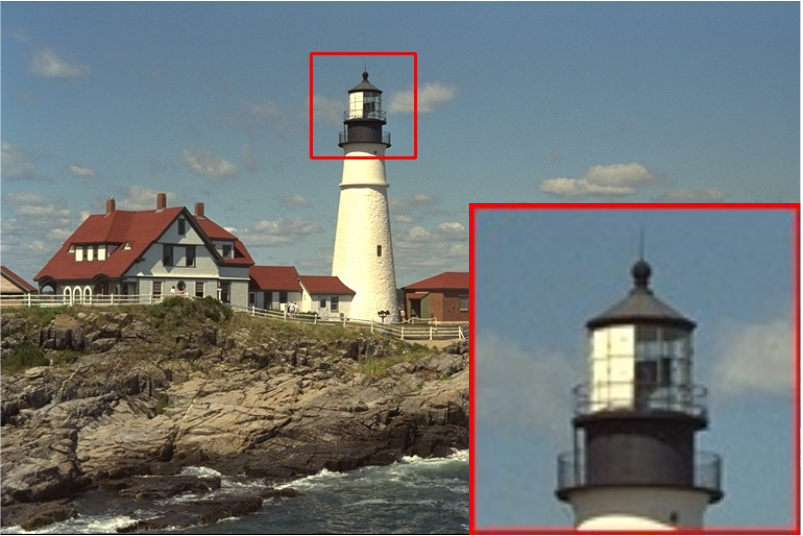}
\includegraphics[width=0.196\textwidth]{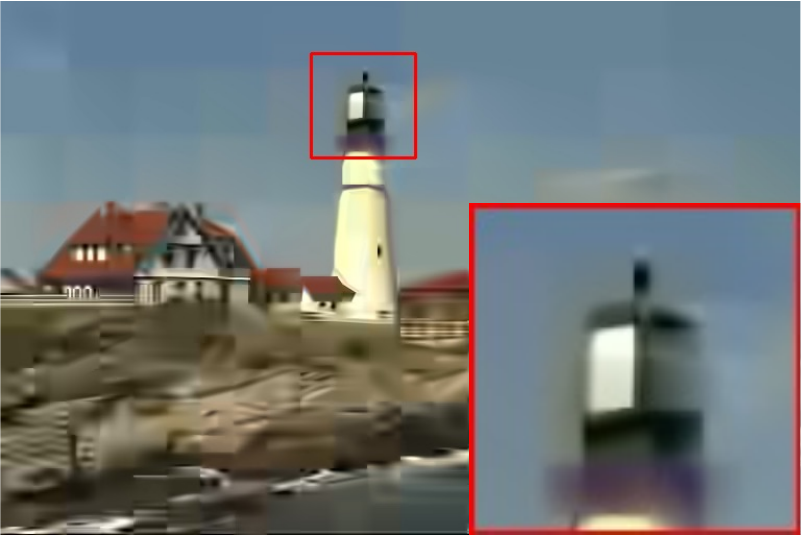}
\includegraphics[width=0.196\textwidth]{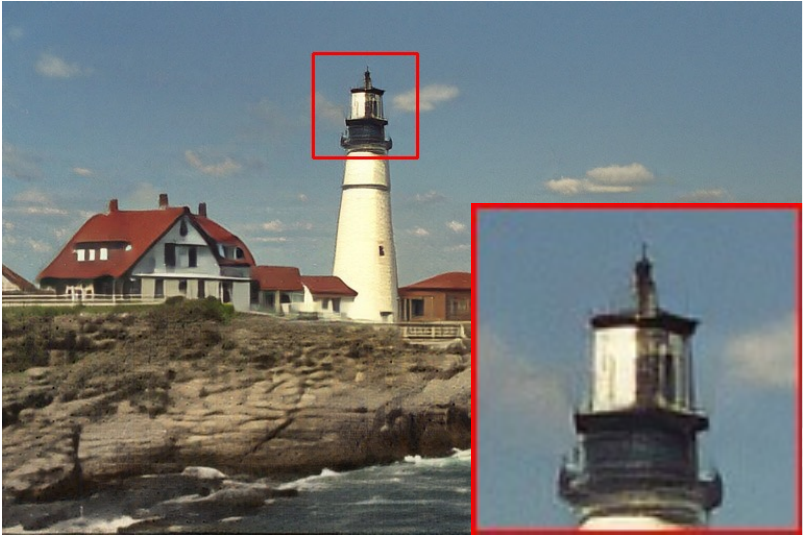}
\includegraphics[width=0.196\textwidth]{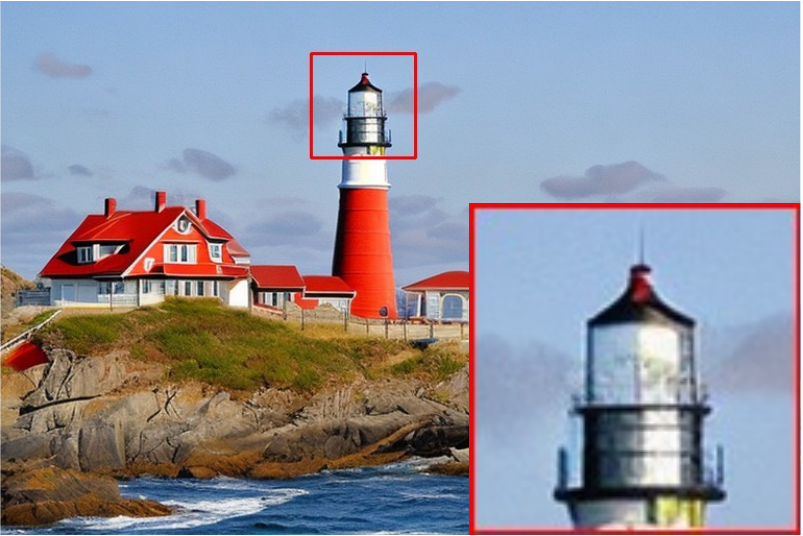}
\includegraphics[width=0.196\textwidth]{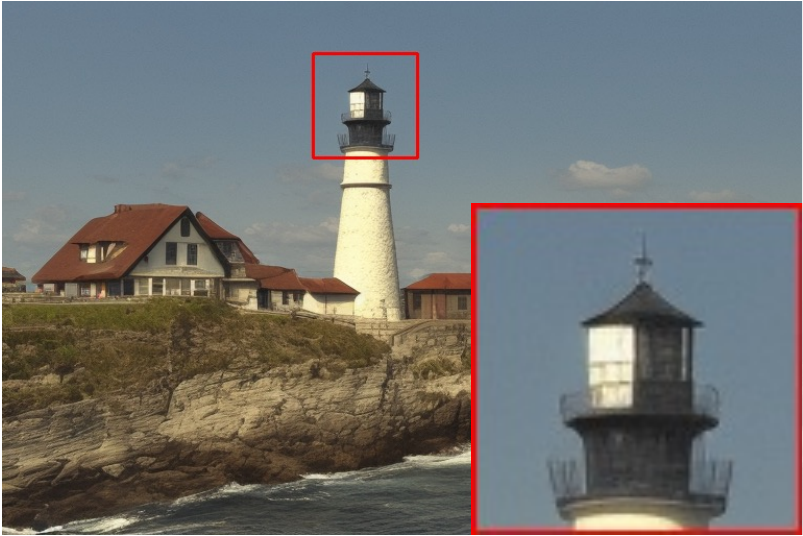} 
\\ \vspace{0.1cm}
\includegraphics[width=0.196\textwidth]{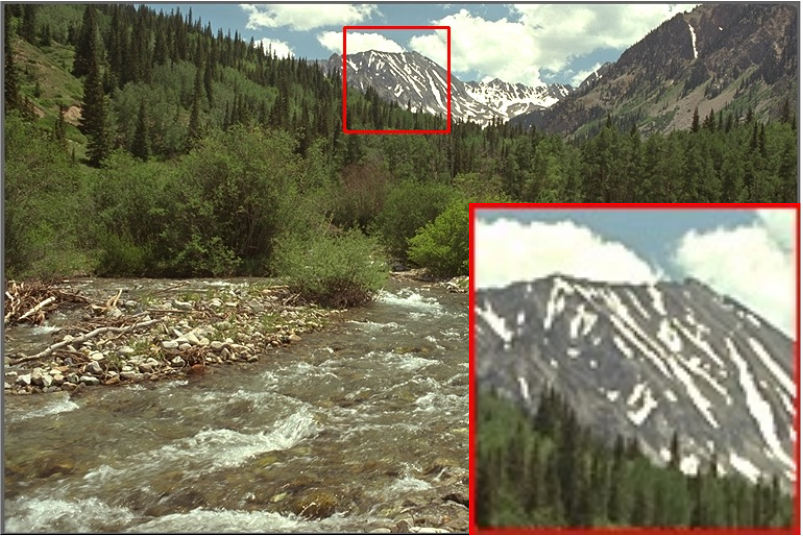}
\includegraphics[width=0.196\textwidth]{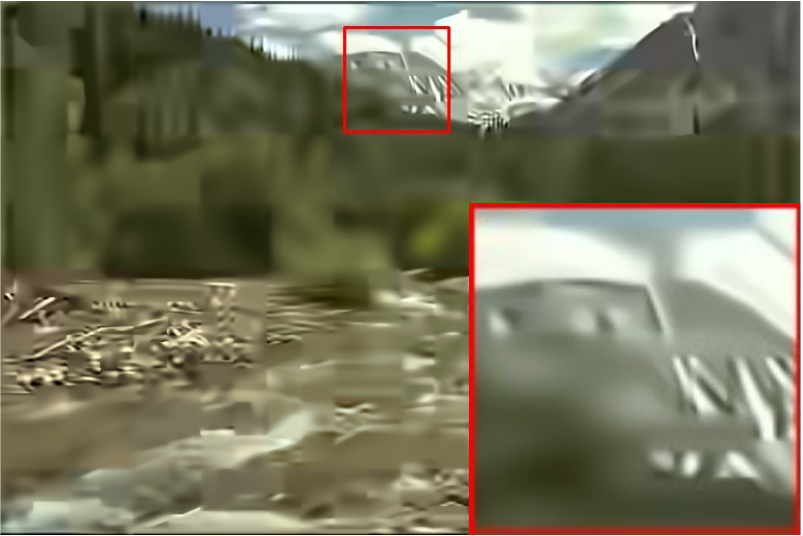}
\includegraphics[width=0.196\textwidth]{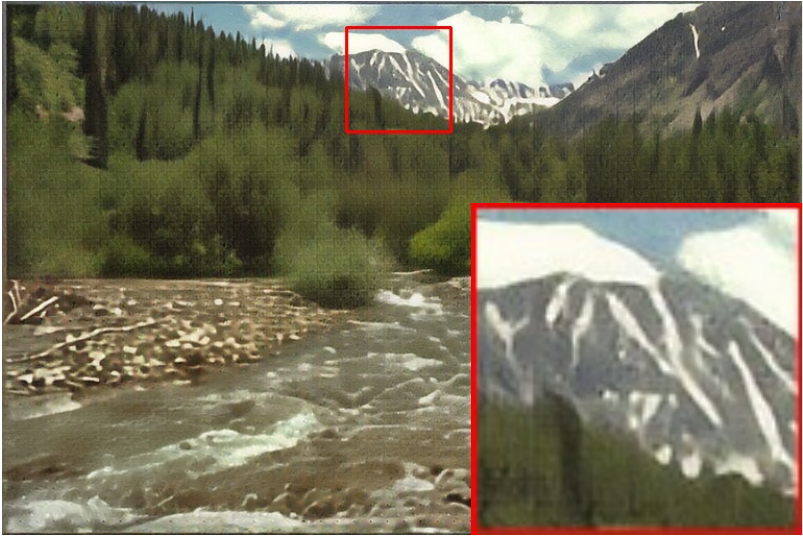}
\includegraphics[width=0.196\textwidth]{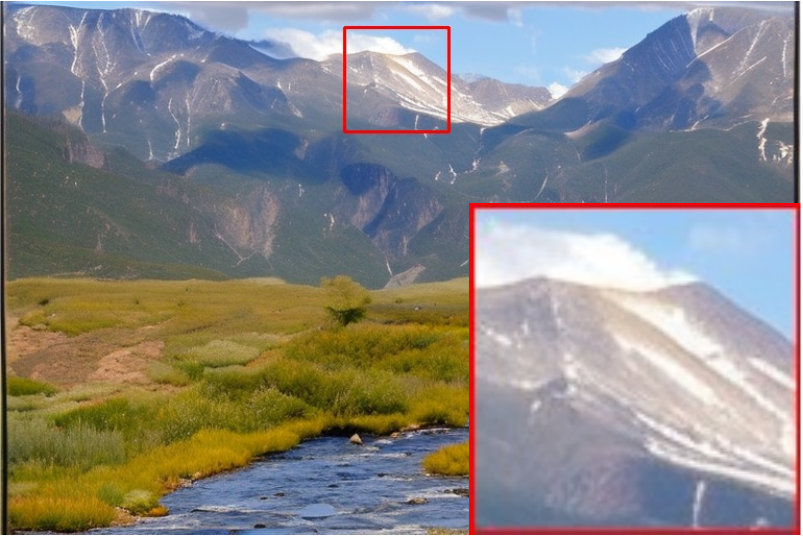}
\includegraphics[width=0.196\textwidth]{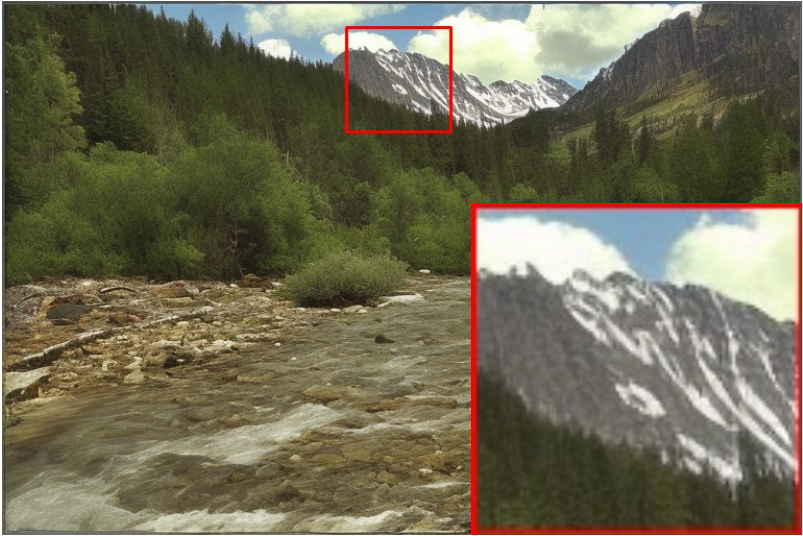}
\\ \vspace{0.1cm}
\includegraphics[width=0.196\textwidth]{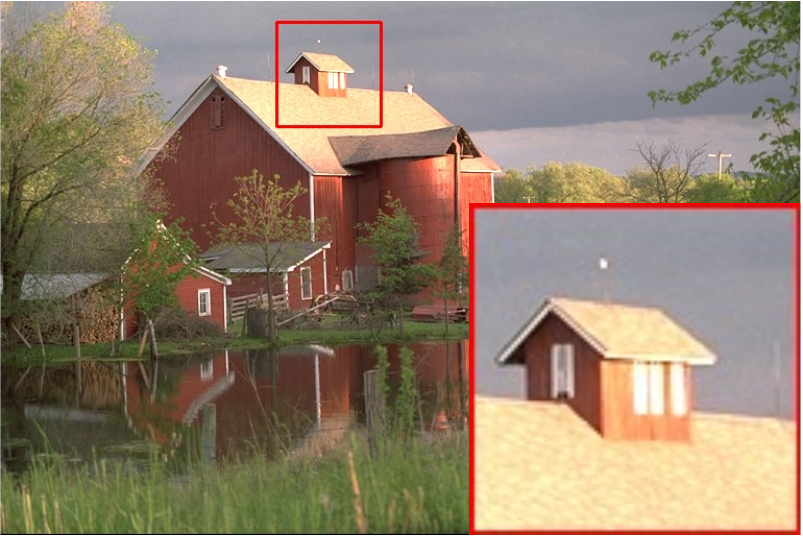}
\includegraphics[width=0.196\textwidth]{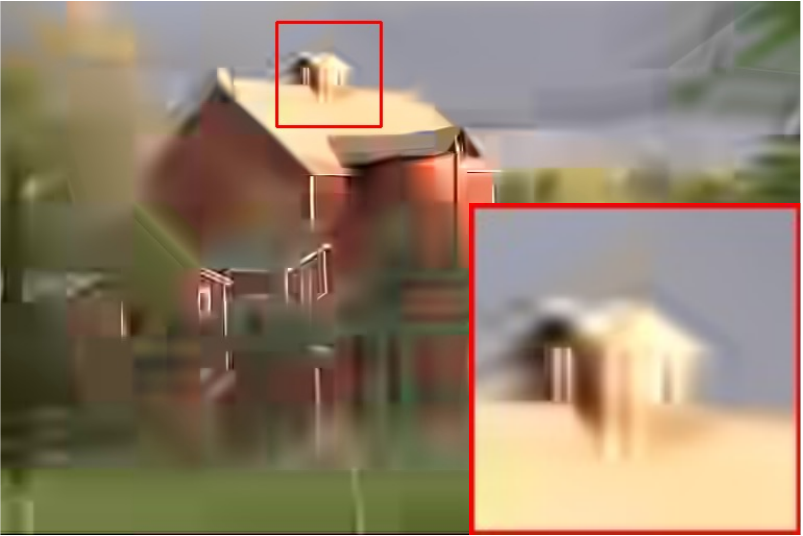}
\includegraphics[width=0.196\textwidth]{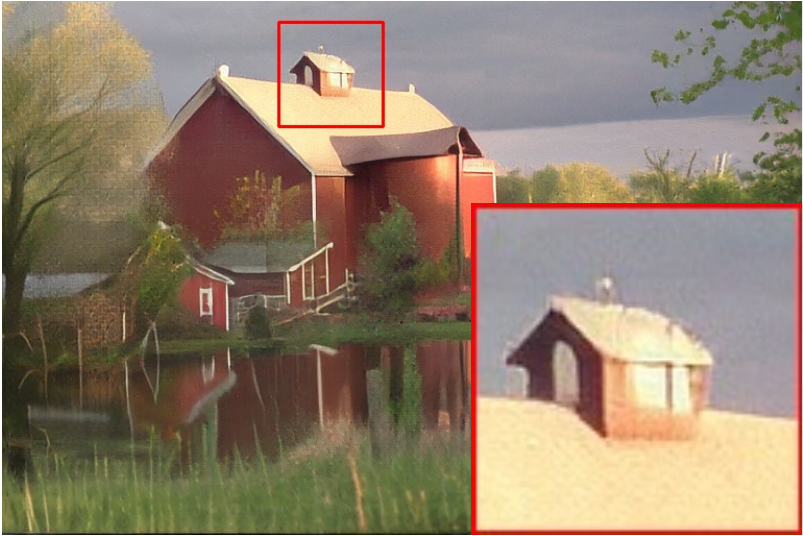}
\includegraphics[width=0.196\textwidth]{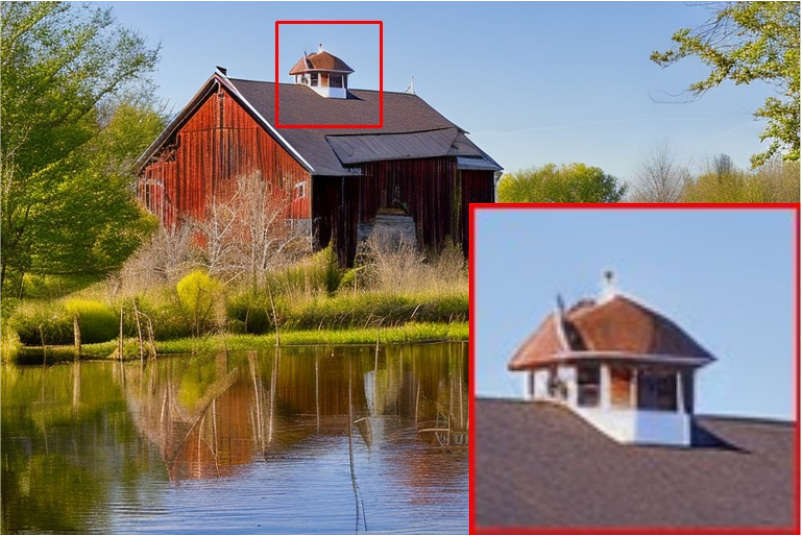}
\includegraphics[width=0.196\textwidth]{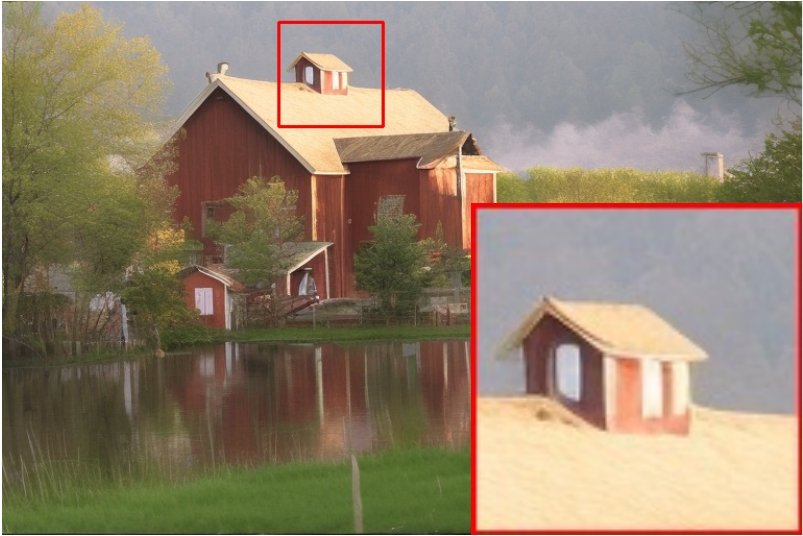}
\caption{Visual examples of the reconstructed results on the Kodak \cite{Kodak} dataset. The proposed DiffEIC produces much better results in terms of perception and fidelity. For example, the small attic is well reconstructed.} 
\label{visual on kodak}
\end{figure*}

\IEEEPARstart{E}{xtreme} image compression, which aims to compress images at bitrates below 0.1 bits per pixel (bpp), is critical in very bandwidth-constrained scenarios, such as satellite communications. Traditional compression standards, such as JPEG2000 \cite{JPEG2000}, BPG \cite{BPG}, and VVC \cite{VVC}, are widely used in practice. However, these algorithms produce severe blocking artifacts at extremely low bitrates due to their block-based processing, see Fig. \ref{visual on kodak}(b).

Learning-based image compression has attracted significant interest and shows great potential to outperform traditional codecs. Based on their optimization objectives, learning-based methods can be roughly categorized into distortion-oriented \cite{TCM, WAttn, Channel-wise, Checkerboard} and perception-oriented \cite{RTAIC, RDP-tradeoff, GANELIC, CompressNet} methods. 
Distortion-oriented methods are optimized for the rate-distortion function, which often leads to unrealistic reconstructions at low bitrates, typically manifested as blurring. 
Perception-oriented methods, on the other hand, aim to optimize the rate-distortion-perception function, leveraging techniques such as adversarial training~\cite{GAN} to improve perceptual quality. While these methods achieve significant improvements in visual quality, they often introduce unpleasant visual artifacts, especially at extremely low bitrates, as shown in Fig. \ref{visual on kodak}(c).

Recently, diffusion models have exhibited impressive generation ability in image and video generation \cite{LDM, ControlNet, SVD}, encouraging researchers to develop various diffusion-based perception-driven compression methods \cite{LCWGD, DIRAC, CDC, PerCo}. For extreme image compression, some works leverage pre-trained text-to-image diffusion models as prior knowledge to achieve realistic reconstructions at extremely low bitrates. For instance, Pan et al. \cite{EGIC} encode images as textual embeddings with extremely low bitrates, using pre-trained text-to-image diffusion models for realistic reconstruction. Lei et al. \cite{Text+Sketch} directly transmit short text prompts and compressed image sketches, employing the pre-trained ControlNet \cite{ControlNet} to produce reconstructions with high perceptual quality and semantic fidelity. However, these methods treat pre-trained text-to-image diffusion models as independent components, which limits their ability to fully exploit the generative capability of pre-trained diffusion models, resulting in reconstruction results that are inconsistent with the original image (see Fig.~\ref{visual on kodak}(d)). Therefore, \emph{how to develop an effective diffusion-based extreme generative compression method is worth further exploration.}

In this work, we develop an end-to-end \textbf{Diff}usion-based \textbf{E}xtreme \textbf{I}mage \textbf{C}ompression (\textbf{DiffEIC}) model that effectively combines compressive variational autoencoders (VAEs) \cite{Hyperprior} with a fixed stable diffusion model.
First, to effectively convey information, we develop a VAE-based latent feature-guided compression module (LFGCM) that can adaptively select information essential for reconstruction, rather than using explicit information, such as text prompts and sketches in \cite{Text+Sketch}, to represent images. {Specifically, this module employs a VAE-based compression method to compress images} and initially decode the compressed information into content variables. {To effectively utilize the knowledge encapsulated in the fixed stable diffusion model, these content variables are expected to align with the diffusion space. However, learning to map images to the diffusion space from scratch is challenging. To address this issue, in the latent feature-guided compression module, we introduce the latent representation of images in the diffusion space as external guidance to correct intermediate features and content variables.}
Second, we introduce a conditional diffusion decoding module (CDDM) to reconstruct images with the guidance of content variables. This module employs the well-trained stable diffusion as a fixed decoder and injects external condition information via a trainable control module. {Leveraging the powerful generative capability of stable diffusion, the proposed DiffEIC can produce realistic reconstructions even at extremely low bitrates. Furthermore, to optimize the model in an end-to-end manner, we design a space alignment loss to force content variables to align with the diffusion space and provide necessary constraints for optimization.} With the help of these mentioned components, the proposed DiffEIC achieves favorable results compared to state-of-the-art approaches, as demonstrated in Fig. \ref{visual on kodak}(e).

In summary, the main contributions of this work are as follows:

1) To the best of our knowledge, we propose the first extreme image compression framework that combines compressive VAEs with pre-trained text-to-image diffusion models in an end-to-end manner.

2) We develop a latent feature-guided compression module to adaptively select information essential for reconstruction. By introducing external guidance, we effectively improve reconstruction fidelity at extremely low bitrates.

3) We propose a conditional diffusion decoding module that fully exploits the powerful diffusion prior contained in the well-trained stable diffusion to facilitate extreme image compression and improve realistic reconstruction.

The remainder of this paper is organized as follows. The related works are summarized in Section \ref{related_work}. The proposed method is described in Section \ref{proposed_method}. The experiment results and analysis are presented in Section \ref{experiments} and Section \ref{analysis}, respectively. Finally, we conclude our work in Section \ref{conclusion}. 

%% file: 03_relatedwork.tex
\section{Related Work}
\label{related_work}
\subsection{Lossy Image Compression}
Lossy image compression plays a crucial role in image storage and transmission. Traditional compression standards, such as BPG \cite{BPG}, HEVC \cite{HEVC} and VVC \cite{VVC}, are widely used in practice. However, they tend to introduce block artifacts due to the lack of consideration of spatial correlation between image blocks.
In recent years, learned image compression has made significant progress and achieved impressive rate-distortion performance \cite{E2EOIC, CCP}. The main success of these methods is attributed to the development of various transform networks and entropy models.
For instance, Liu et al. \cite{NLattention} introduce a non-local attention module to improve transform networks. In \cite{INC}, He et al. employ invertible neural networks (INNs) to mitigate the information loss problem. Zhu et al. \cite{ViTLIC} construct nonlinear transforms using swin-transformers, achieving superior compression performance compared to CNNs-based transforms. In \cite{CDC}, Yang et al. innovatively use conditional diffusion models as decoders. 
Furthermore, several methods \cite{Channel-wise, Checkerboard, Autoregressive} enhance performance by improving entropy models. For example, Minnen et al. \cite{Autoregressive} combine hierarchical priors with autoregressive models to reduce spatial redundancy within latent features. In \cite{ELIC}, He et al. assume the redundancy in spatial dimension and channel dimension is orthogonal and propose a multi-dimension entropy model. Qian et al. \cite{Entroformer} utilize a transformer to enable entropy models to capture long-range dependencies. {Guo et al.~\cite{CCP} explore capturing the dependencies along both the spatial and channel dimensions by using the causal global contextual prediction.}

\subsection{Extreme Image Compression}
In some practical scenarios, such as satellite communications, the bandwidth is too narrow to transmit the images or videos. To overcome this dilemma, extreme image compression towards low bitrates (e.g., below 0.1 bpp) is urgently needed. 
Several algorithms \cite{GANELIC, CompressNet, FCEIC, HiFiC, MS-ILLM} leverage generative adversarial networks (GANs) for realistic reconstructions and bit savings. In \cite{GANELIC}, Agustsson et al. incorporate a multi-scale discriminator to synthesize details that cannot be stored at extremely low bitrates. Mentzer et al. \cite{HiFiC} explore normalization layers, generator and discriminator architectures, training strategies, as well as perceptual losses, achieving visually pleasing reconstructions at low bitrates. However, these approaches suffer from the unstable training of GANs and inevitably introduce unpleasant visual artifacts. 

Some approaches use prior knowledge to achieve extreme image compression. Yue et al. \cite{Cloud-Based_Image_Coding} describe input images based on the down-sampled version and handcrafted features, and use these descriptions to reconstuct the images from a large-scale image database. Their method can achieve impressive compression performance when the large-scale image database contains images that are highly correlated with the input images. Benefiting from the bijective and information-lossless property of invertible neural networks (INNs), Gao et al. leverage INNs to mitigate the significant information loss in extreme image compression \cite{INVEIC}.
Wei et al. employ invertible and generative priors to achieve extreme compression by rescaling images with extreme scaling factors (i.e., 16$\times$ and 32$\times$) \cite{VQIR}.  
In \cite{underwater}, Li et al. employ physical priors (i.e., attenuation coefficient and ambient light) and the semantic prior for extreme underwater image compression, which may not generalize well to natural images with different scenarios. Jiang et al. utilize text descriptions as prior to guide image compression for better compression performance \cite{TGIC}.

Inspired by the tremendous success of diffusion models in image generation, some methods \cite{EGIC, Text+Sketch} use more powerful pre-trained text-to-image diffusion models as prior knowledge. In \cite{EGIC}, Pan et al. encode images into short text embeddings and then generate high-quality images with pre-trained text-to-image diffusion models by feeding the text embeddings. Lei et al. \cite{Text+Sketch} directly compress the short text prompts and binary contour sketches on the encoded side, and then use them as input to the pre-trained text-to-image diffusion model for reconstruction on the decoded side. However, these diffusion-based methods treat pre-trained text-to-image diffusion models as independent components, which limits their ability to fully exploit the generative capability of pre-trained diffusion models.

In this work, we propose DiffEIC, a framework that efficiently incorporates compressive VAEs with pre-trained text-to-image diffusion models in an end-to-end manner. Leveraging the nonlinear capability of compressive VAEs and the powerful generative capability of pre-trained text-to-image diffusion models, our DiffEIC achieves both high perceptual quality and high-fidelity image reconstruction at extremely low bitrates.

\subsection{Diffusion Models}
Inspired by non-equilibrium statistical physics \cite{DM}, diffusion models convert real data distributions into simple, known distributions (e.g., Gaussian) through a gradual process of adding random noise, known as the diffusion process. Subsequently, they learn to reverse this diffusion process and construct desired data samples from noise (i.e., the reverse process).
Denoising diffusion implicit models (DDPM) \cite{DDPM} improves upon the original diffusion model and has profoundly influenced subsequent research. The latent diffusion model (LDM) \cite{LDM} significantly reduces computational costs by performing diffusion and reverse steps in the latent space. Stable diffusion is a widely used large-scale implementation of LDM. Owing to their flexibility, tractability, and superior generative capability, diffusion models have achieved remarkable success in various vision tasks. 

Due to the complexity of the diffusion process, training diffusion models from scratch is computationally demanding and time-consuming. To address this problem, some algorithms \cite{ControlNet, T2I-Adapter, DiffBIR} introduce additional trainable networks to inject external conditions into fixed, pre-trained diffusion models. This strategy simplifies the exhaustive training from scratch while maintaining the robust capability of pre-trained diffusion models.
In \cite{ControlNet}, Zhang et al. employ pre-trained text-to-image diffusion models (e.g., stable diffusion) as a strong backbone with fixed parameters and reuse their encoding layers for controllable image generation. Similarly, Mou et al. \cite{T2I-Adapter} introduce lightweight T2I-Adapters to provide extra guidance for pre-trained text-to-image diffusion models. In \cite{DiffBIR}, Lin et al. use the latent representation of coarse restored images as conditions to help the pre-trained diffusion models generate clean results.
We note that the main success of these algorithms on image generation and restoration is due to the use of pre-trained diffusion models. The robust generative capability of such models motivates us to explore effective approaches for extreme image compression at low bitrates.

%% file: 04_algorithm.tex
\section{Methodology}
\label{proposed_method}
\begin{figure*}[htbp]\scriptsize
\centering
\includegraphics[width=0.98\textwidth]{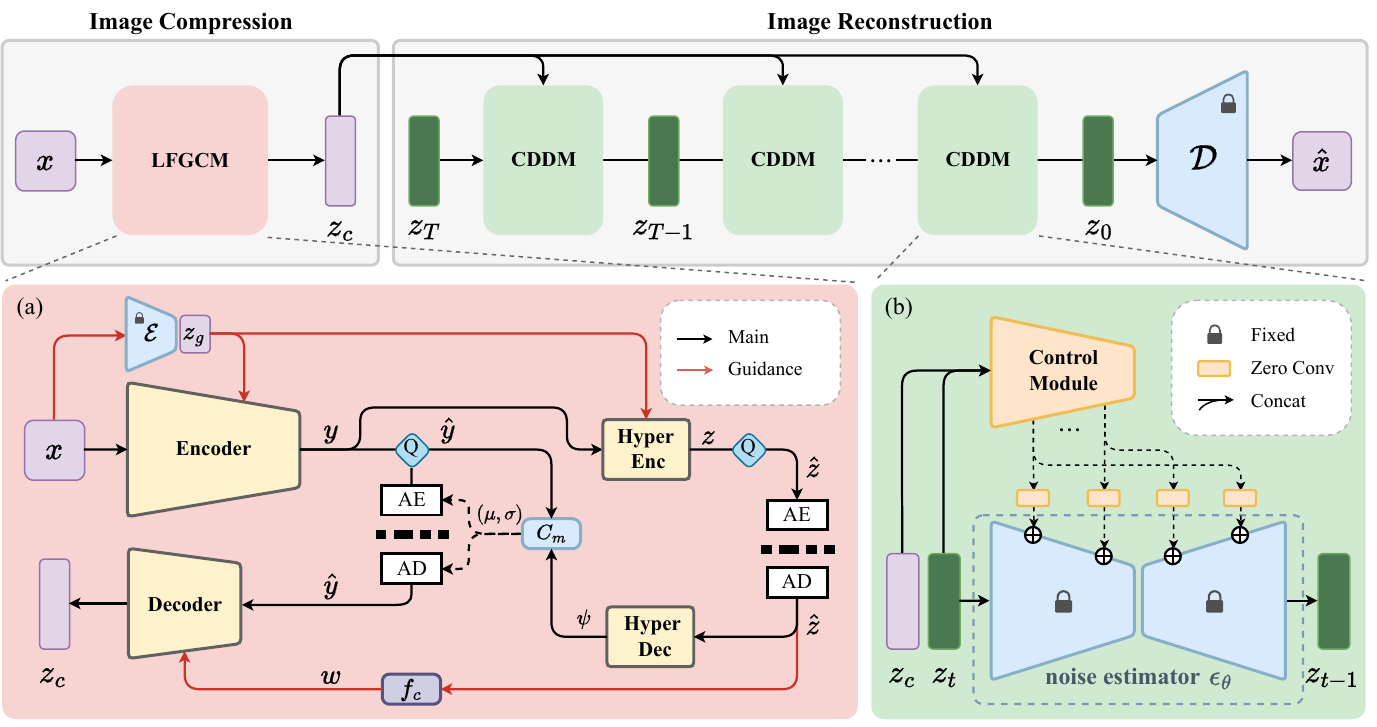}
\caption{The two-stage pipeline of the proposed DiffEIC. \textbf{Image Compression:} Initially, we leverage the VAE-based latent feature-guided compression module (LFGCM) to adaptively select information essential for reconstruction and obtain $z_c$. \textbf{Image Reconstruction:} We leverage the conditional diffusion decoding module (CDDM) for realistic image reconstruction and obtain $\hat x$. The CDDM contains a trainable control module and a fixed noise estimator. Note that the control module and noise estimator are connected with zero convolutions (zero-initialized convolution layers).
}
\label{arch_framework}
\end{figure*}

In this section, we propose DiffEIC for extreme image compression. As shown in Fig. \ref{arch_framework}, the proposed DiffEIC consists of two primary stages: image compression and image reconstruction. Specifically, the former stage aims to compress images and generate content-related variables. The latter stage is designed for decoding the content variables into reconstructed images. Furthermore, a space alignment loss is introduced to force content variables to align with the diffusion space and provide necessary constraints for optimization.

\subsection{Image Compression with Compressive VAEs}\label{IC}
As shown in Fig. \ref{arch_framework}(a), we propose a latent feature-guided compression module (LFGCM) based on compressive VAEs \cite{Hyperprior}. This module leverages an additional guidance branch that utilizes the latent representation of images in the diffusion space to correct intermediate features and content variables. The encoding process, decoding process, and network details of LFGCM are introduced below.
\subsubsection{Encoding Process}
Given an input image $x$, we first obtain external guidance $z_g$ with stable diffusion's encoder $\mathcal{E}$ as follows:
\begin{equation}
    z_g = \mathcal{E}(x).
\end{equation}
Then $z_g$ is used to guide the extraction of the latent representation $y$ and the side information $z$, sequentially, which can be expressed as:
\begin{equation}
    y = \mathcal{N}_{e}(x, z_g), \  z = \mathcal{N}_{he}(y, z_g),
 \end{equation}
where $\mathcal{N}_{e}$ denotes the encoder network and $\mathcal{N}_{he}$ denotes the hyper-encoder network. Then we apply a hyper-decoder to draw a parameter $\psi$ from the quantized side information $\hat{z}$:
\begin{equation}
    \hat{z} = \mathcal{Q}(z), \ \psi = \mathcal{N}_{hd}(\hat{z}),
\end{equation}
where $\mathcal{N}_{hd}$ denotes the hyper-decoder and $\mathcal{Q}(\cdot)$ denotes the quantization operation, i.e., adding uniform noise during training and performing rounding operation during inference. Finally, the context model $\mathcal{C}_{m}$ uses $\psi$ and the quantized latent representation $\hat{y}=\mathcal{Q}(y)$ to predict the Gaussian entropy parameters $(\mu, \sigma)$ for approximating the distribution of $\hat{y}$.

\subsubsection{Decoding process}
Given the quantized $\hat{y}$ and $\hat{z}$, we first use a information extraction network $f_c$ to extract a representation $w$ from $\hat{z}$, which can be expressed as:
\begin{equation}
    w = f_c(\hat{z}).
\end{equation}
The external guidance information, originally contained in $z_g$, is captured in $w$. This effectively compensates for the unavailability of $z_g$ during the decoding process. Instead of directly reconstructing the original input image, we initially decode $\hat{y}$ into a content variable $z_c$:
\begin{equation}
    z_c = \mathcal{N}_{d}(\hat{y}, w),
\end{equation}
where $\mathcal{N}_{d}$ denotes the decoder network. The content variable $z_c$ is further decoded in the subsequent image reconstruction stage using diffusion prior.

\subsubsection{Network Details}
\begin{figure*}[t]\scriptsize
\centering
\includegraphics[width=0.98\textwidth]{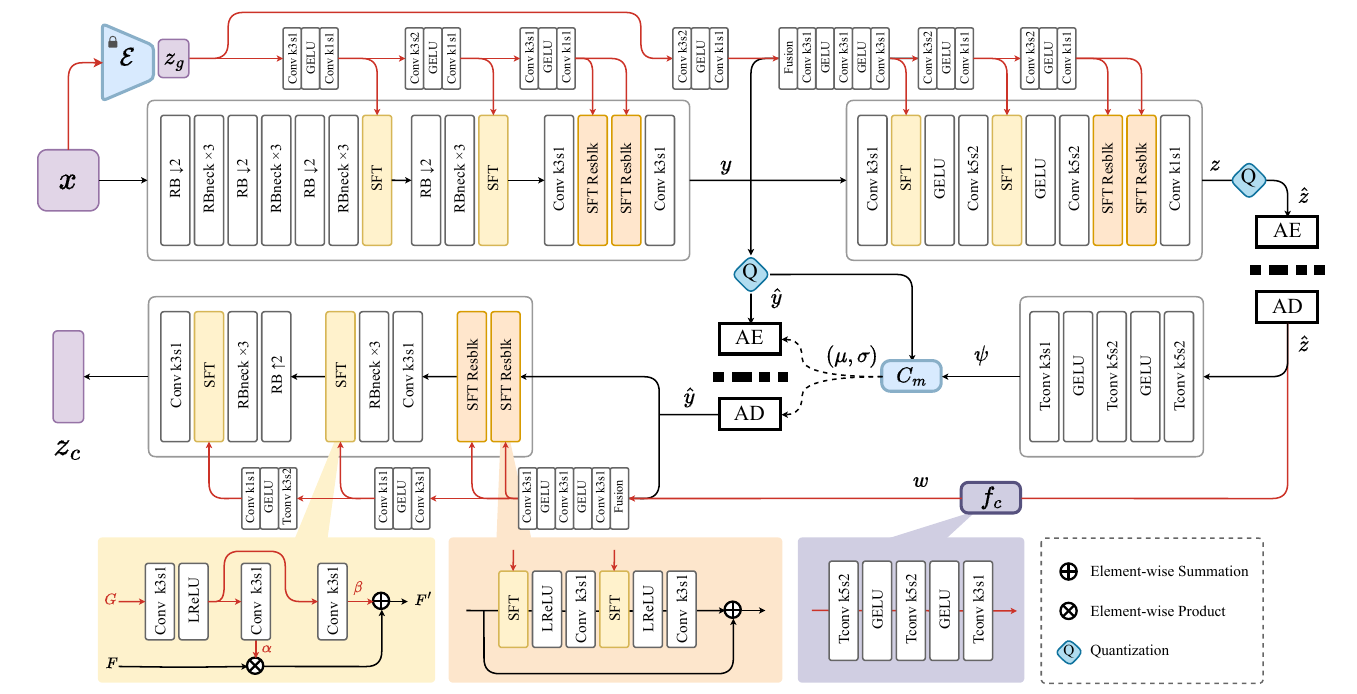}
\caption{The architecture of the proposed LFGCM. $Conv\ k3s1$ denotes convolution with $3\times 3$ filters and stride 1. $Tconv\ k3s1$ denotes transposed convolution with $3\times 3$ filters and stride 1. $RB$ denotes residual block \cite{resnet}. $RBneck$ denotes residual bottleneck block \cite{resnet}. $LReLU$ denotes the LeakyReLU function. $AE$ and $AD$ denote arithmetic encoder and decoder, respectively. $C_m$ denotes context model. $Fusion$ denotes the fusion method. The black and red arrows denote main and guidance flow, respectively.}
\label{LFGCM_arch}
\end{figure*}

Fig. \ref{LFGCM_arch} illustrates the network architecture of LFGCM. The information extraction network $f_c$ has the same structure as the hyper-decoder $\mathcal{N}_{hd}$, and we adopt context model $\mathcal{C}_{m}$ proposed by He et al. \cite{ELIC}. The guidance components (the elements denoted by red arrows, $SFT$ and $SFT \ Resblk$) first use a series of convolutions to resize the external feature $G$ to the appropriate dimensions. Then the SFT layers \cite{SFT} are employed to inject the network with external guidance information. Specifically, given an external feature $G$ and an intermediate feature map $F$, a pair of affine transformation parameters (i.e., $\alpha$ for scaling and $\beta$ for shifting) is generated as follows:
\begin{equation}
   \alpha, \beta = \Phi_\theta (G),
\end{equation}
where $\Phi_\theta$ denotes a stack of convolutions. Then the tuned feature map $F'$ can be generated by:
\begin{equation}
    F'=SFT(F,G)= \alpha \otimes F + \beta,
\end{equation}
where $\otimes$ denotes element-wise product.

\subsection{Image Reconstruction with Diffusion Prior}
As shown in Fig. \ref{arch_framework}(b), we propose a conditional diffusion decoding module (CDDM) to {reconstruct images with the guidance of content variables}. {To maintain the generative capability of stable diffusion, we keep it fixed and employ a small control module to inject content information.} In this section, we introduce stable diffusion and the proposed CDDM sequentially.

\subsubsection{Stable Diffusion}
Stable diffusion first employs an encoder $\mathcal{E}$ to encode an image $x$ into a latent representation $z_0= \mathcal{E}(x)$. Then $z_0$ is progressively corrupted by adding Gaussian noise through a Markov chain. The intensity of the added noise at each step is controlled by a default noise schedule $\beta_t$. This process can be expressed as follows:
\begin{equation}
    z_t = \sqrt{\bar{\alpha_t}}z_{0} + \sqrt{1-\bar{\alpha_t}}\epsilon ,\  t=1,2,\cdots,T,
\end{equation}
where $\epsilon \sim \mathcal{N}(0,\textbf{I})$ is a sample from a standard Gaussian distribution, $\alpha_t = 1-\beta_t$ and $\bar{\alpha_t}=\prod_{i=1}^t \alpha_i$. The corrupted representation $z_t$ approaches a Gaussian distribution as $t$ increases. To iteratively convert $z_T$ back to $z_0$, a noise estimator $\epsilon_\theta$ with U-Net \cite{U-net} architecture is learned to predict the added noise $\epsilon$ at each time step $t$:
\begin{equation}
    \mathcal{L}_{sd} = \mathbb{E}_{z_0,c,t,\epsilon}\Vert \epsilon - \epsilon_\theta(z_t,c,t)\Vert^2,
    \label{l_sd}
\end{equation}
where $c$ denotes control conditions such as text prompts and images. After completing the iterative denoising process, a decoder $\mathcal{D}$ is used to map $z_0$ back into pixel space.

\subsubsection{Conditional Diffusion Decoding Module}
{The CDDM is designed to leverage the powerful generative capability of fixed stable diffusion to reconstruct image $x$ with realistic details at extremely low bitrates. Inspired by ControlNet \cite{ControlNet},} we introduce a control module to inject content information contained in $z_c$ into the denoising process. This control module has the same encoder and middle block architecture as the noise estimator $\epsilon_\theta$. Notably, we reduce the channel number of the control module to 20\% of the original{, which results in a slight performance decrease but significantly enhances inference speed (see Section \ref{ablation_of_cn}).} In addition, we increase the channel number of the first convolution layer to 8 to accommodate the concatenated input of the content variable $z_c$ and the latent noise $z_t$. Through the control module, we obtain a series of conditional features that contain content information and align with the internal knowledge of stable diffusion. These conditional features are then added to the encoder and decoder of the noise estimator $\epsilon_\theta$ using 1$\times$1 convolutions. Leveraging the powerful generative capability encapsulated in pre-trained stable diffusion, we can obtain a high perceptual quality reconstruction $\hat{x}$ even at extremely low bitrates.

\begin{figure*}[htbp]\scriptsize
\centering
\includegraphics[width=0.96\textwidth]{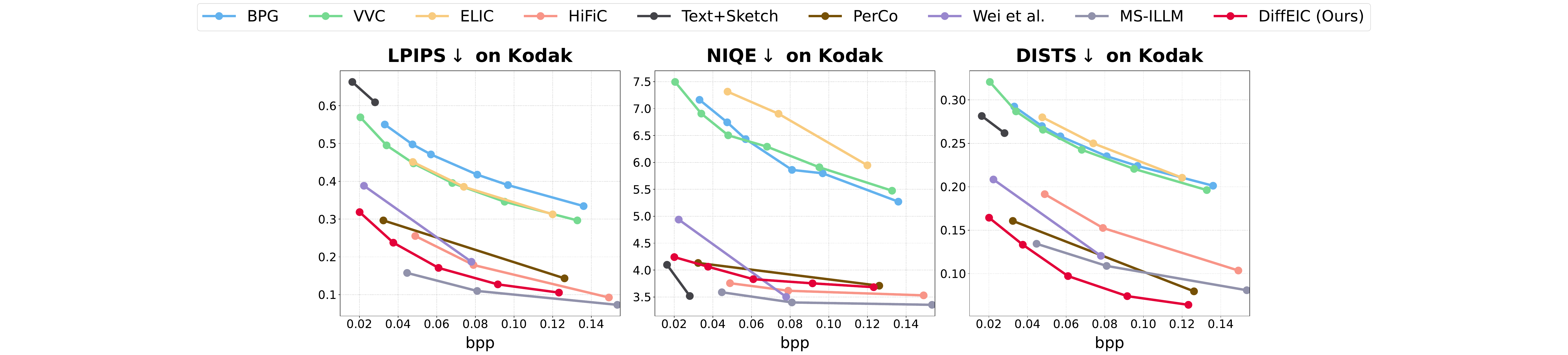} \\ 
\vspace{0.1cm}
\includegraphics[width=0.96\textwidth]{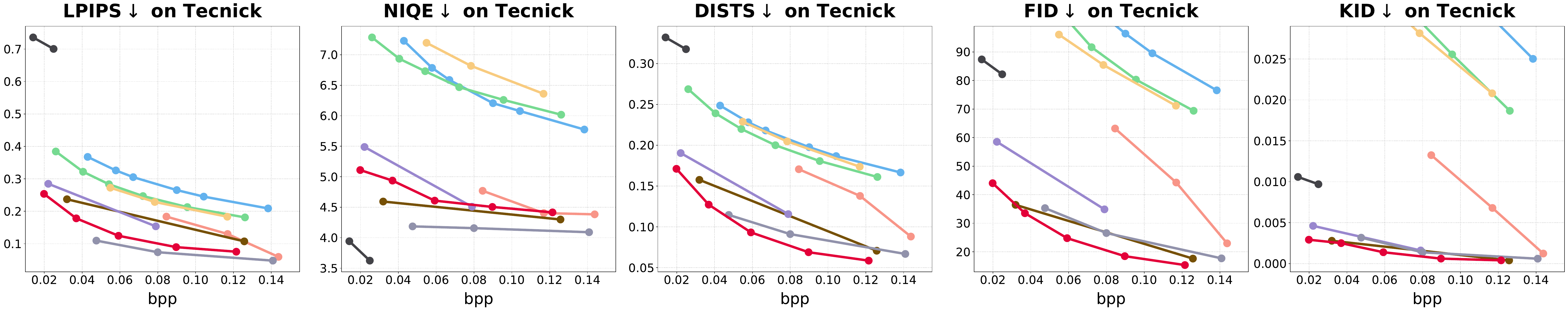} \\ 
\vspace{0.1cm}
\includegraphics[width=0.96\textwidth]{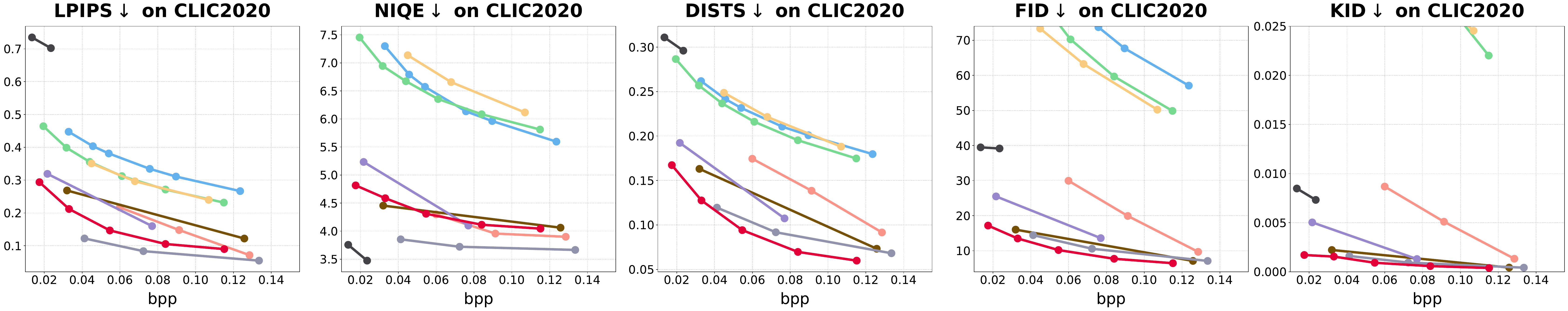} \\ 
\caption{Quantitative comparisons with state-of-the-art methods in terms of perceptual quality (LPIPS$\downarrow$ / NIQE$\downarrow$ / {DISTS$\downarrow$} / FID$\downarrow$ / KID$\downarrow$) on the Kodak \cite{Kodak}, Tecnick \cite{Tecnick}, and CLIC2020 \cite{CLIC2020} datasets.}
\label{R-D-P}
\end{figure*}

\begin{figure*}[htbp]\scriptsize
\centering
\includegraphics[width=0.98\textwidth]{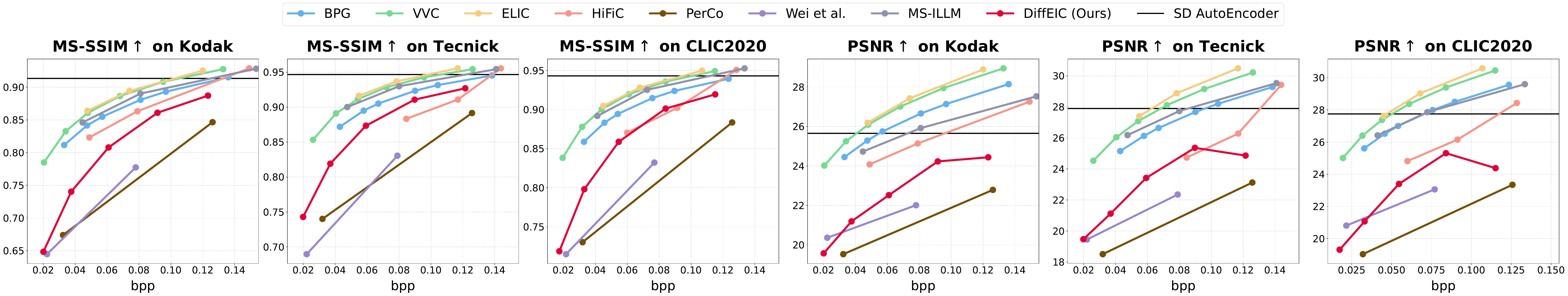} \\ 
\caption{Quantitative comparisons with state-of-the-art methods in terms of pixel fidelity (MS-SSIM$\uparrow$ / PSNR$\uparrow$) on the Kodak \cite{Kodak}, Tecnick \cite{Tecnick}, and CLIC2020 \cite{CLIC2020} datasets.}
\label{R-D}
\end{figure*}

\subsection{Model Objectives}
\subsubsection{Noise Estimation Loss}
Due to the external condition $z_c$ introduced by proposed CDDM, Eq. (\ref{l_sd}) is modified as:
\begin{equation}
    \mathcal{L}_{ne} = \mathbb{E}_{z_0,c,t,\epsilon,z_c}\Vert \epsilon - \epsilon_\theta(z_t,c,t,z_c)\Vert^2,
\end{equation}
where text prompt $c$ is set to empty.

\subsubsection{Rate loss}
We employ the rate loss $\mathcal{L}_{rate}$ to optimize the rate performance as:
\begin{equation}
    \mathcal{L}_{rate} = R(\hat{y}) + R(\hat{z}),
\end{equation}
where $R(\cdot)$ denotes the bitrate.

\subsubsection{Space Alignment Loss}
{As the noise estimation loss is unable to provide effective constraints for LFGCM,} we design a space alignment loss to force the content variables to align with the diffusion space, providing necessary constraints for optimization:
\begin{equation}
    \mathcal{L}_{sa} = \Vert z_c - \mathcal{E}(x) \Vert ^2 .
\end{equation}

In summary, the total loss of DiffEIC is defined as:
\begin{equation}
    \mathcal{L}_{total} = \lambda \mathcal{L}_{rate} + \lambda_{sa} \mathcal{L}_{sa} + \lambda_{ne} \mathcal{L}_{ne},
    \label{loss}
\end{equation}
where $\lambda_{sa}$ and $\lambda_{ne}$ denote the weights for space alignment loss and noise estimation loss, respectively. $\lambda$ is used to achieve a trade-off between rate and reconstruction quality.

%% file: 05_experiments.tex
\section{Experiments}
\label{experiments}

\begin{figure*}[htbp]\scriptsize
\centering
\makebox[0.12\textwidth]{(a) \textbf{Original}}
\makebox[0.12\textwidth]{(b) \textbf{ELIC}}
\makebox[0.12\textwidth]{(c) \textbf{HiFiC}}
\makebox[0.12\textwidth]{{(d) \textbf{MS-ILLM}}}
\makebox[0.12\textwidth]{(e) \textbf{Text+Sketch}}
\makebox[0.12\textwidth]{(f) \textbf{PerCo}}
\makebox[0.12\textwidth]{(g) \textbf{Wei et al.}}
\makebox[0.12\textwidth]{(h) \textbf{DiffEIC (Ours)}}
\\ \vspace{0.1cm}
\includegraphics[width=0.12\textwidth]{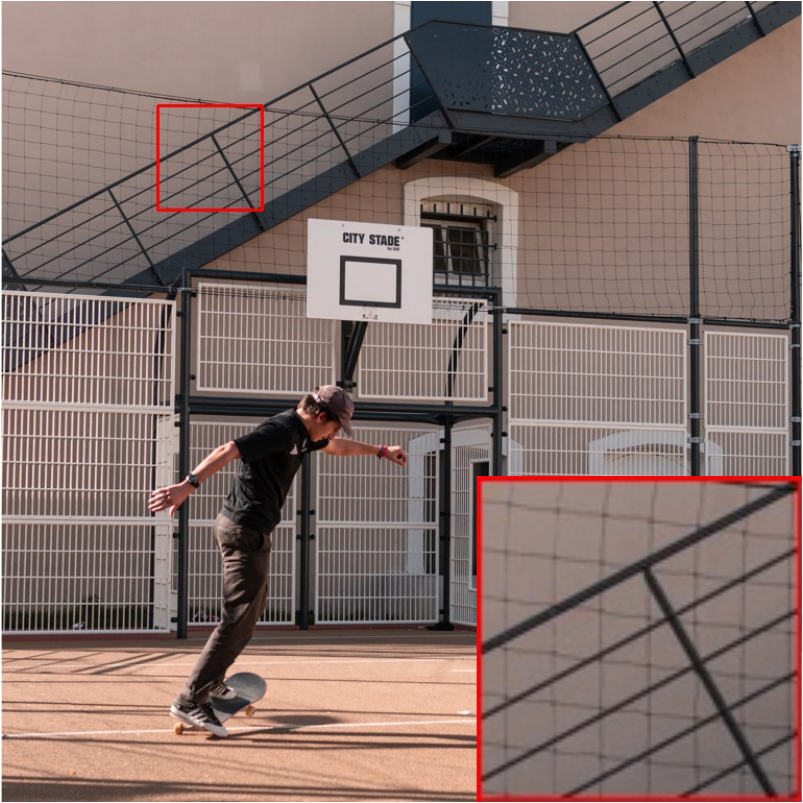}
\includegraphics[width=0.12\textwidth]{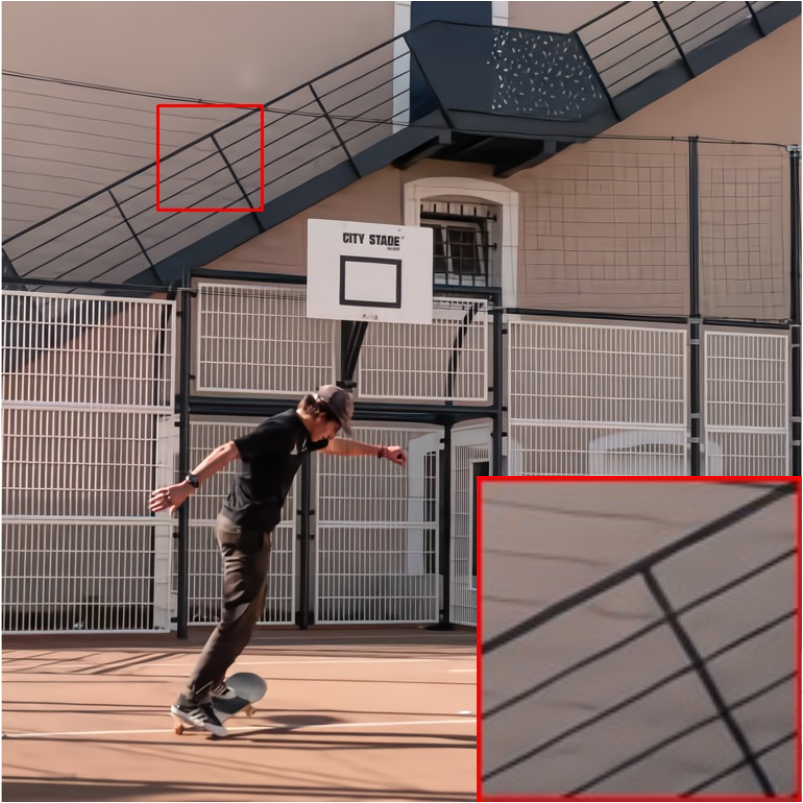}
\includegraphics[width=0.12\textwidth]{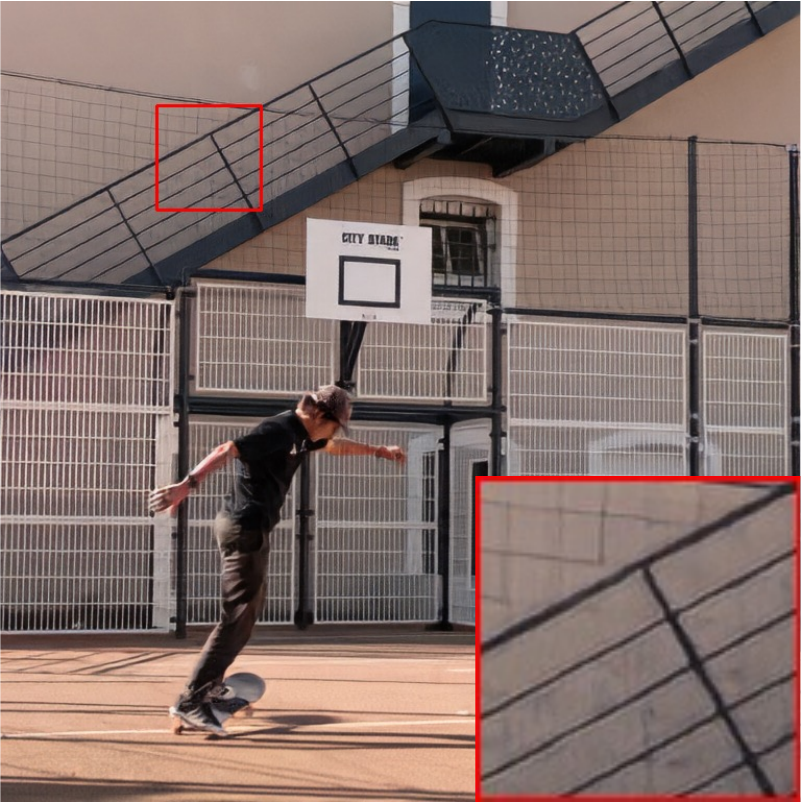}
\includegraphics[width=0.12\textwidth]{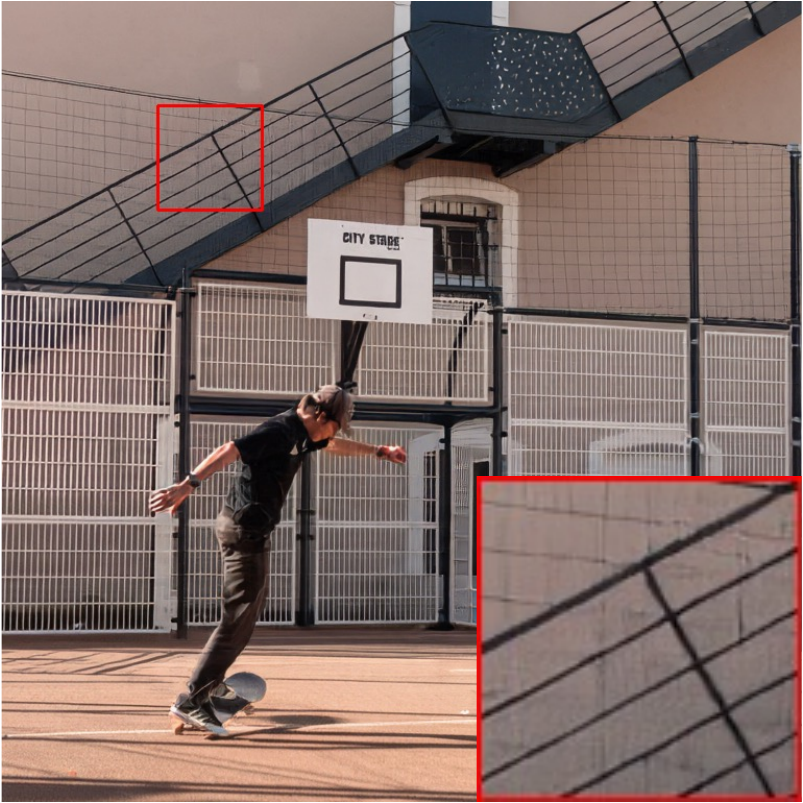}
\includegraphics[width=0.12\textwidth]{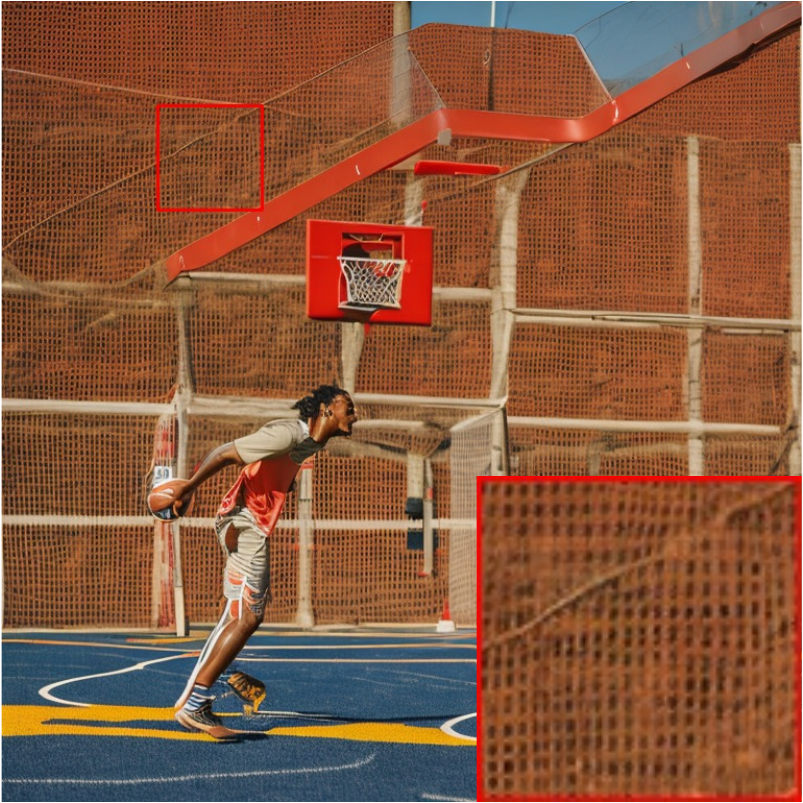}
\includegraphics[width=0.12\textwidth]{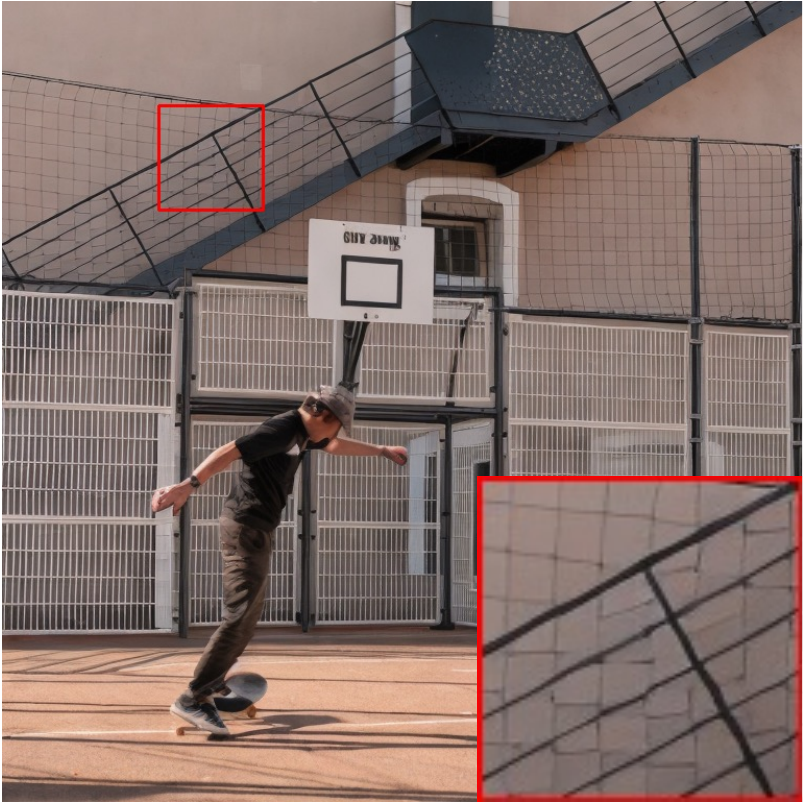}
\includegraphics[width=0.12\textwidth]{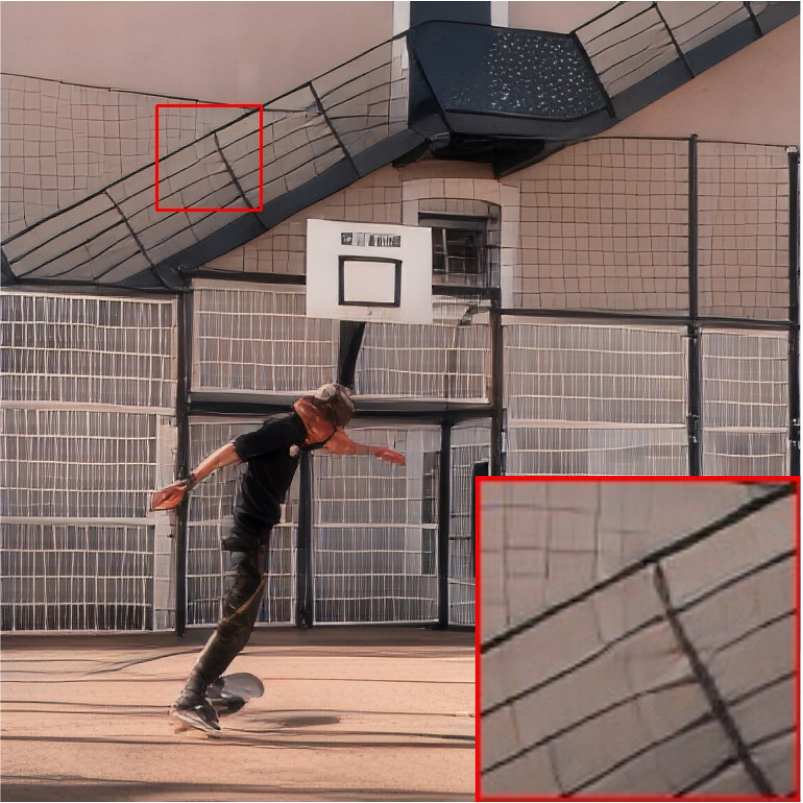} 
\includegraphics[width=0.12\textwidth]{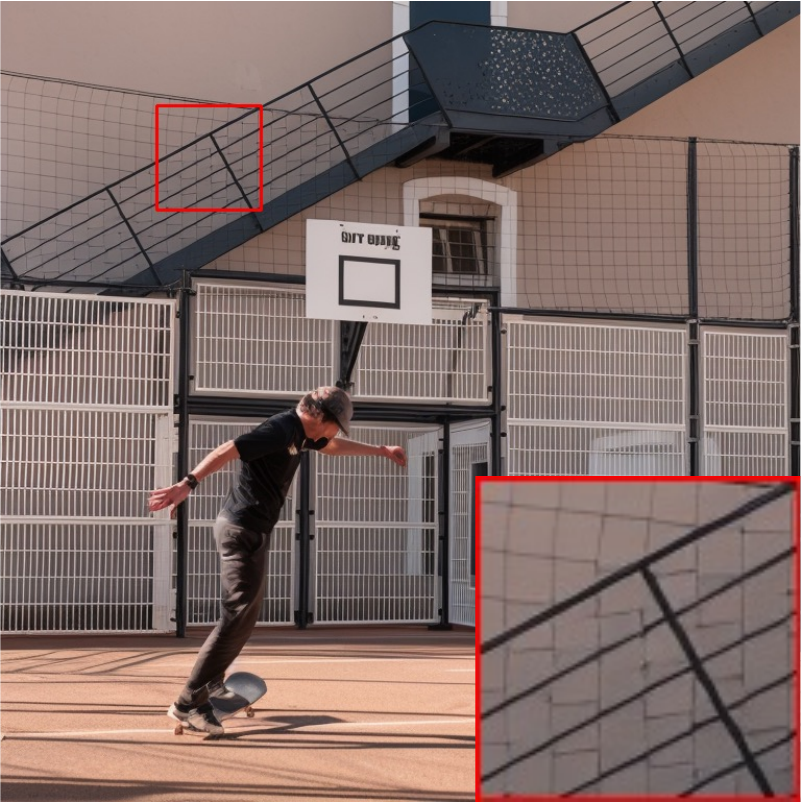} 
\\ \vspace{0.1cm}
\makebox[0.12\textwidth]{\textbf{bpp / DISTS$\downarrow$}}
\makebox[0.12\textwidth]{\textbf{0.1784 / 0.1296}}
\makebox[0.12\textwidth]{\textbf{0.1901 / 0.1178}}
\makebox[0.12\textwidth]{\textbf{0.1662 / 0.0741}}
\makebox[0.12\textwidth]{\textbf{0.0246 / 0.3135}}
\makebox[0.12\textwidth]{\textbf{0.1258 / 0.0674}}
\makebox[0.12\textwidth]{\textbf{0.0792 / 0.1431}}
\makebox[0.12\textwidth]{\textbf{0.0844 / 0.0658}}
\\ \vspace{0.1cm}
\includegraphics[width=0.12\textwidth]{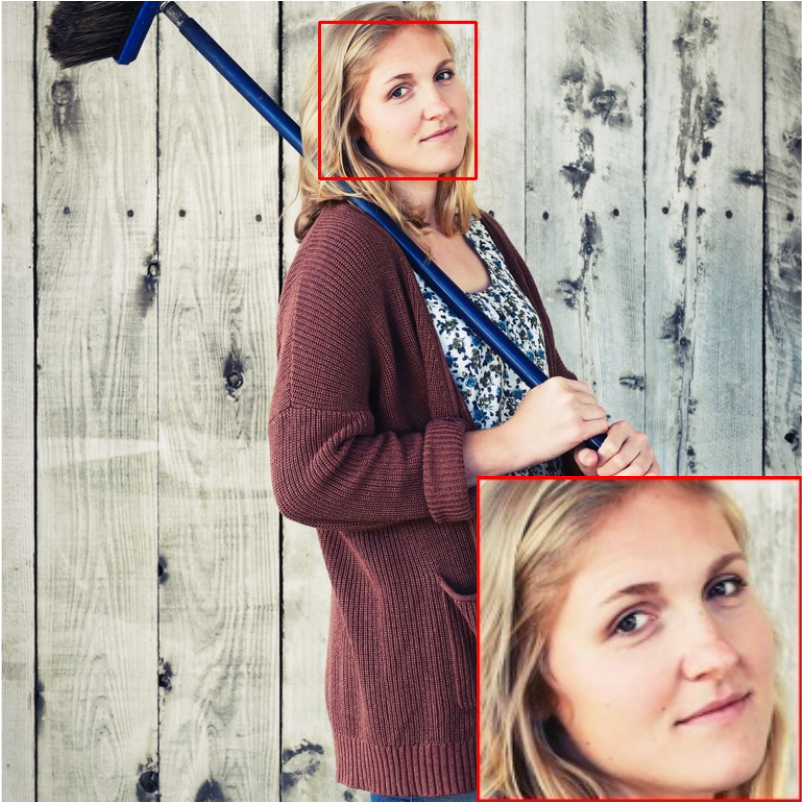}
\includegraphics[width=0.12\textwidth]{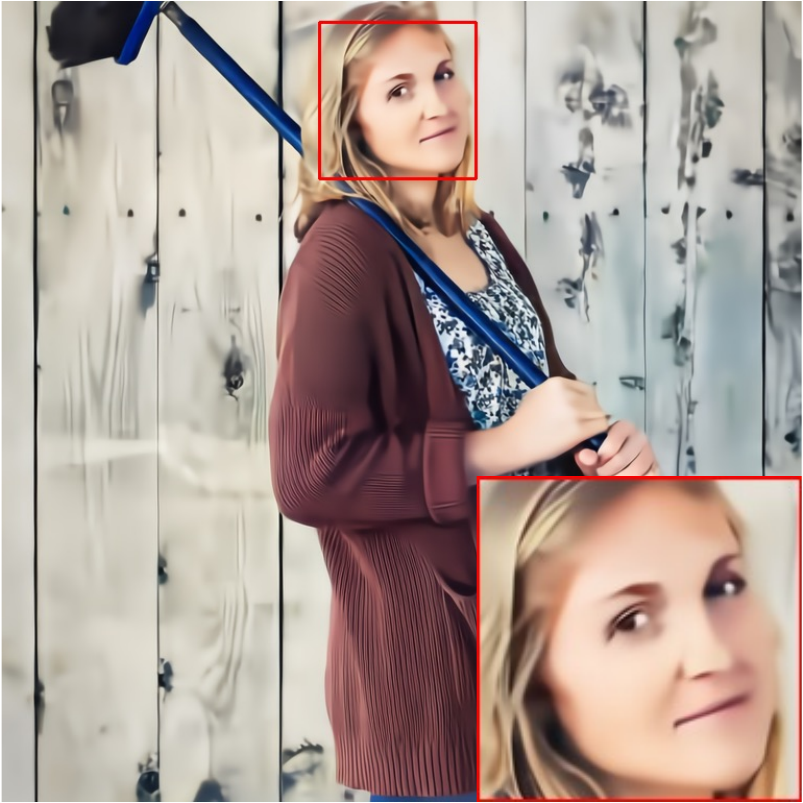}
\includegraphics[width=0.12\textwidth]{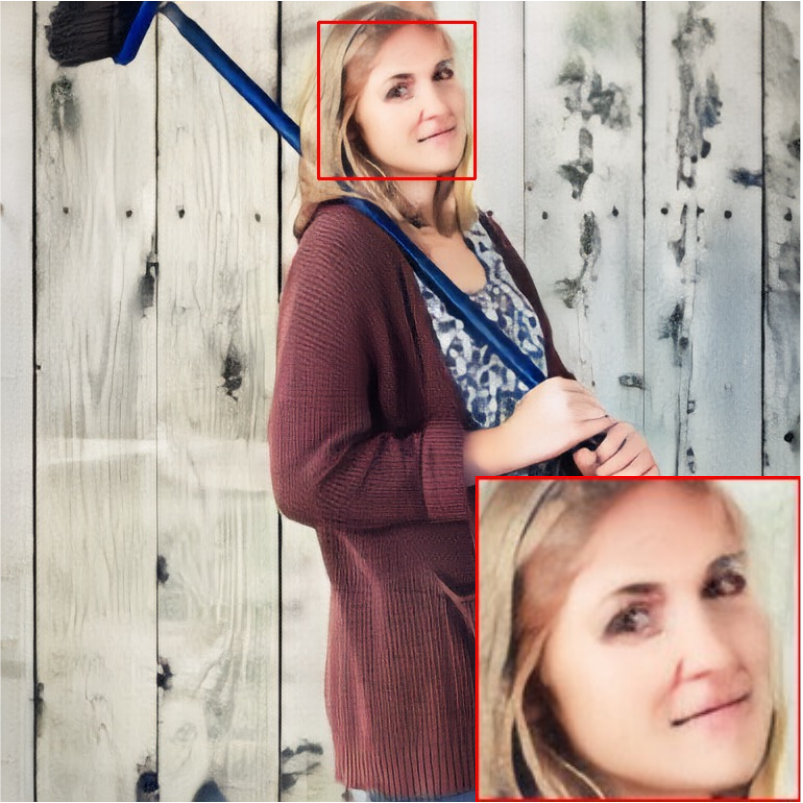}
\includegraphics[width=0.12\textwidth]{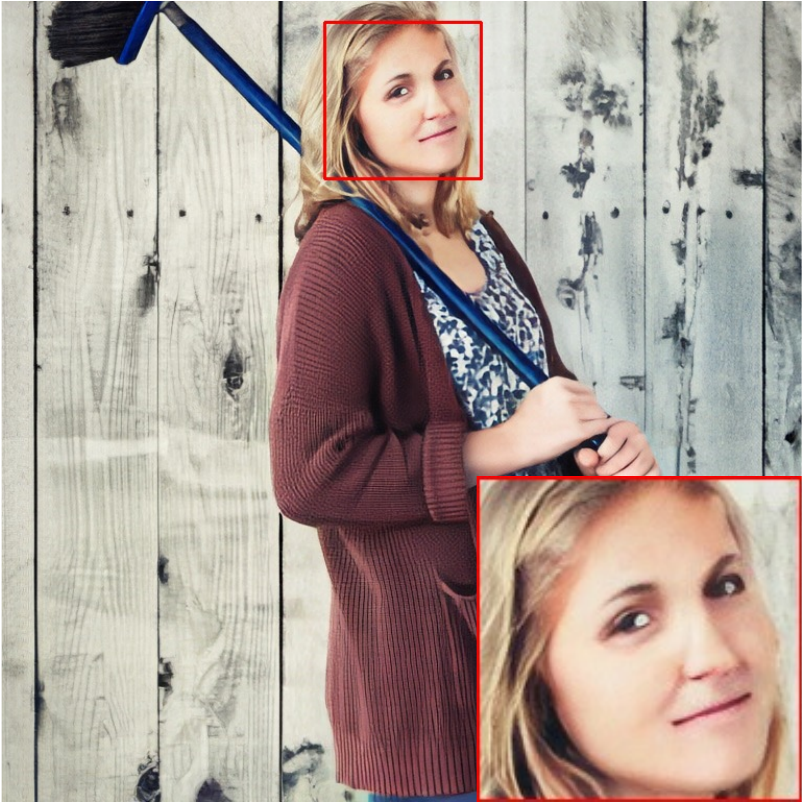}
\includegraphics[width=0.12\textwidth]{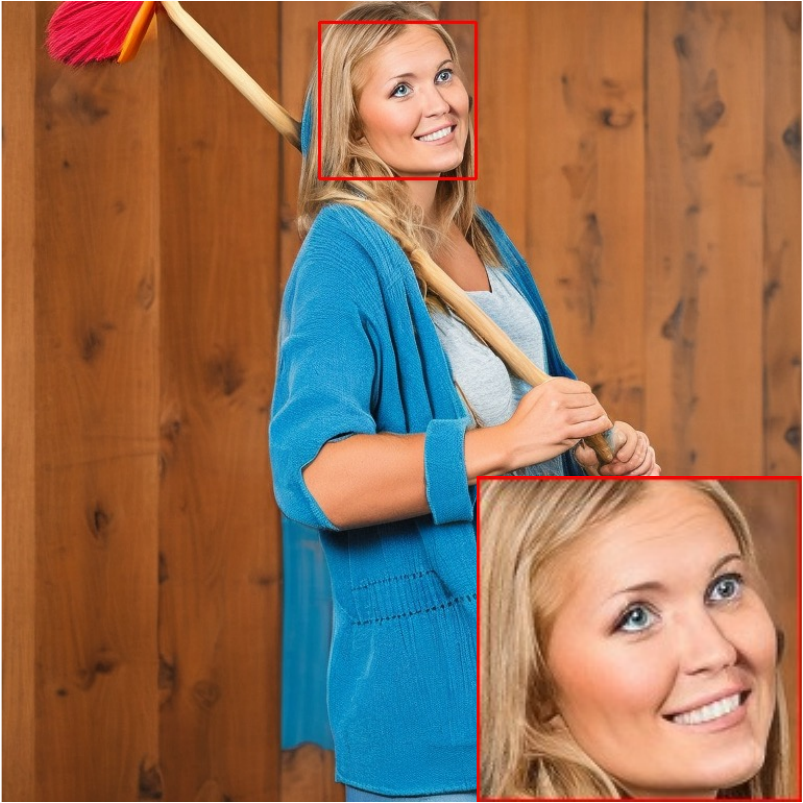}
\includegraphics[width=0.12\textwidth]{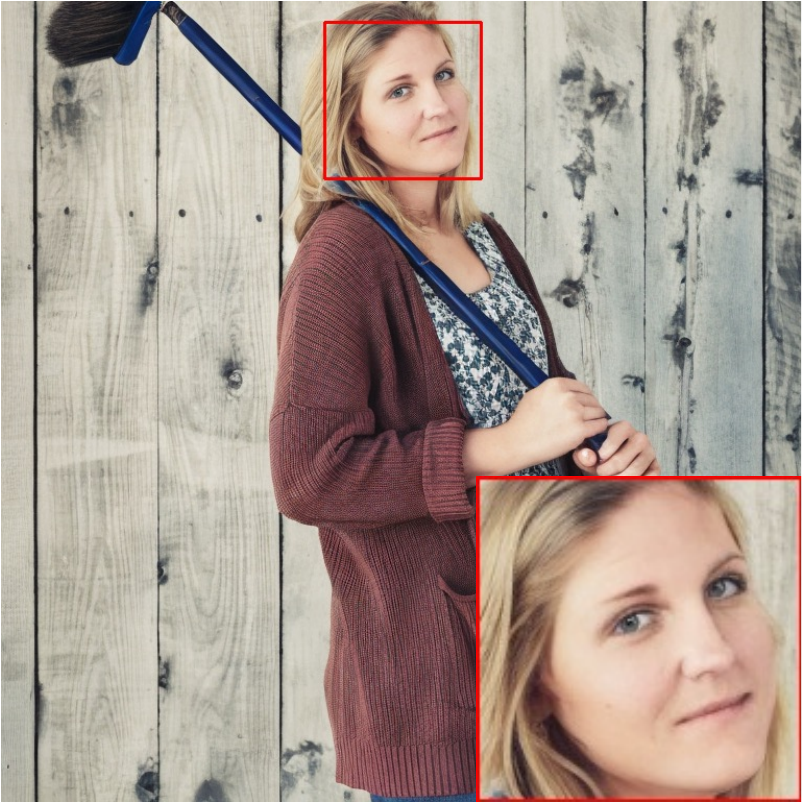}
\includegraphics[width=0.12\textwidth]{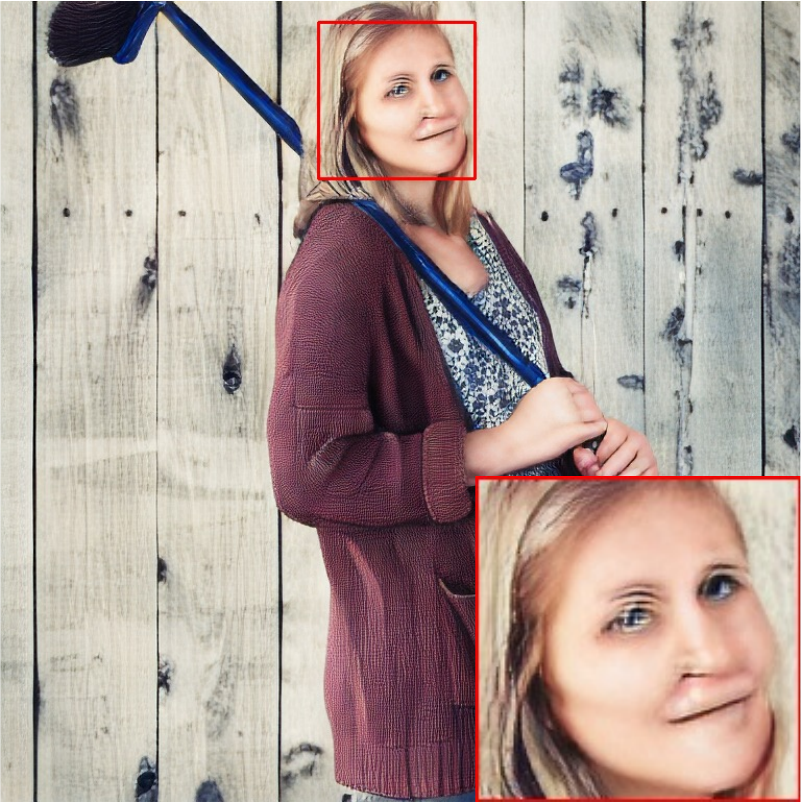} 
\includegraphics[width=0.12\textwidth]{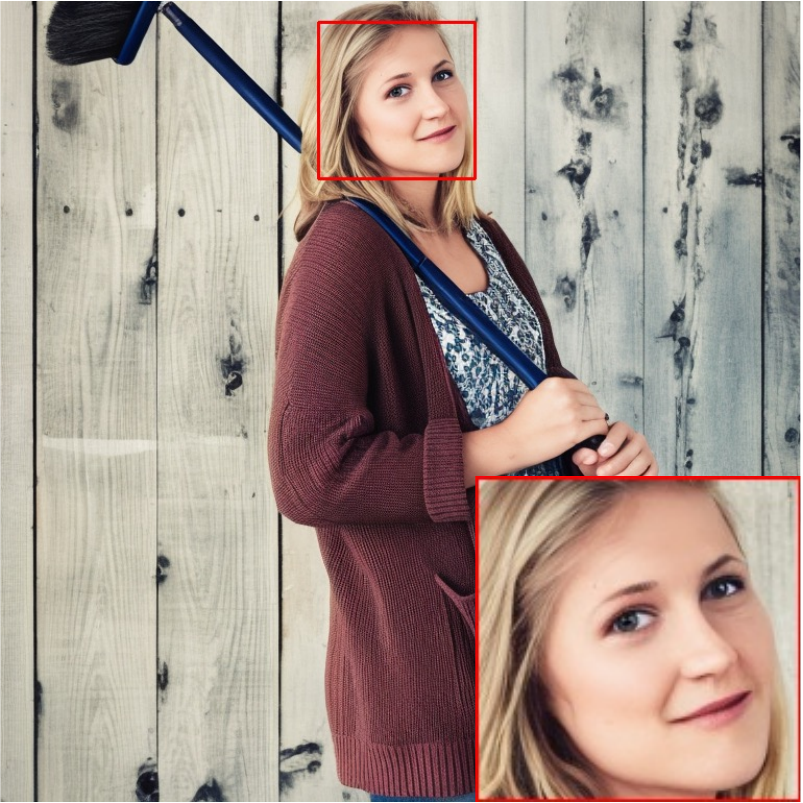} 
\\ \vspace{0.1cm}
\makebox[0.12\textwidth]{\textbf{bpp / DISTS$\downarrow$}}
\makebox[0.12\textwidth]{\textbf{0.0947 / 0.2136}}
\makebox[0.12\textwidth]{\textbf{0.0800 / 0.1193}}
\makebox[0.12\textwidth]{\textbf{0.0893 / 0.0766}}
\makebox[0.12\textwidth]{\textbf{0.0228 / 0.2661}}
\makebox[0.12\textwidth]{\textbf{0.1258 / 0.0704}}
\makebox[0.12\textwidth]{\textbf{0.0838 / 0.0891}}
\makebox[0.12\textwidth]{\textbf{0.0664 / 0.0775}}
\\ \vspace{0.1cm}
\includegraphics[width=0.12\textwidth]{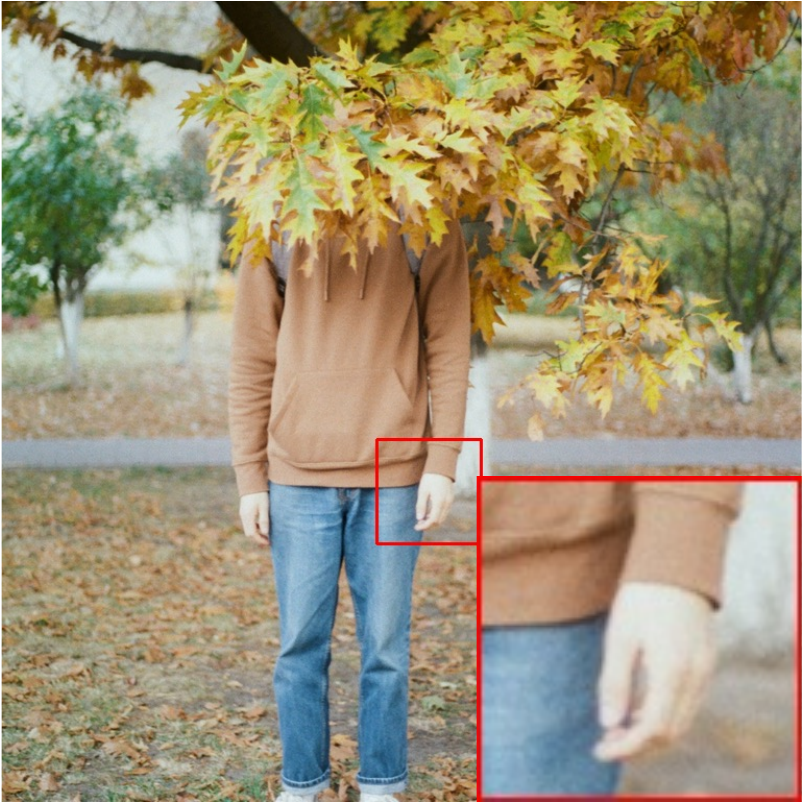}
\includegraphics[width=0.12\textwidth]{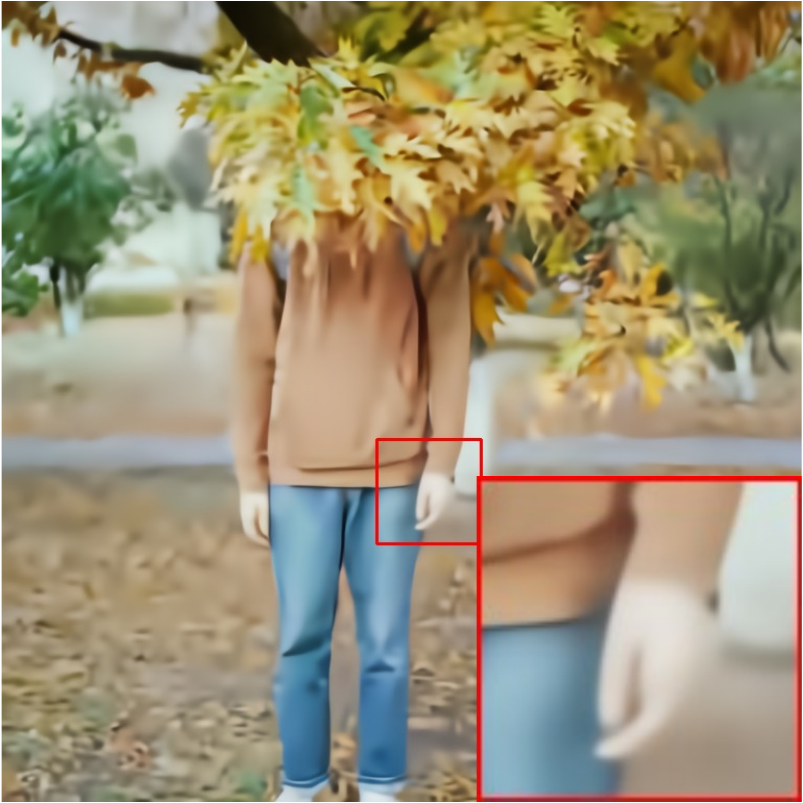}
\includegraphics[width=0.12\textwidth]{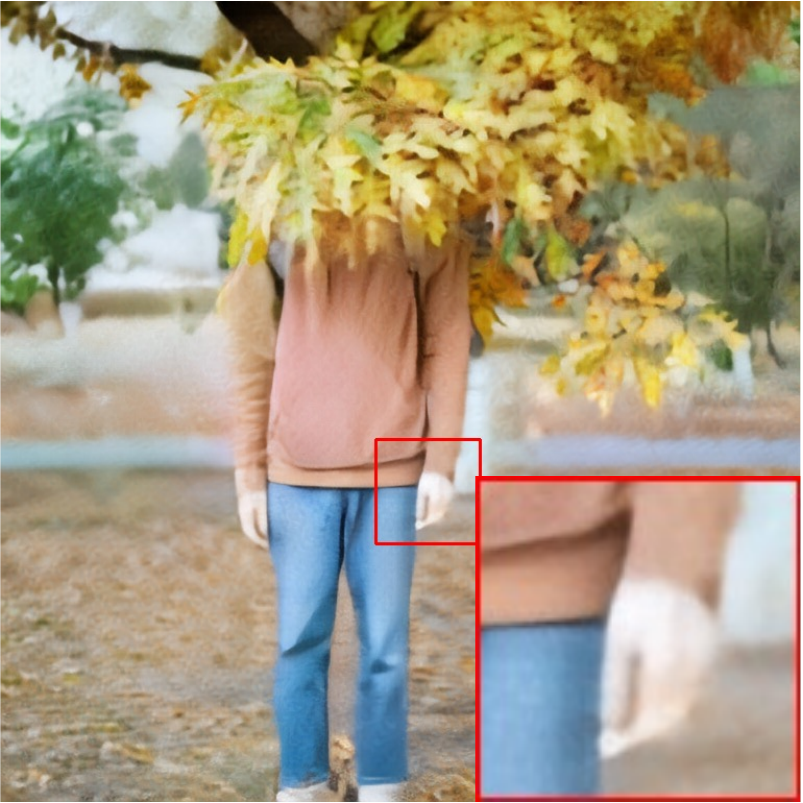}
\includegraphics[width=0.12\textwidth]{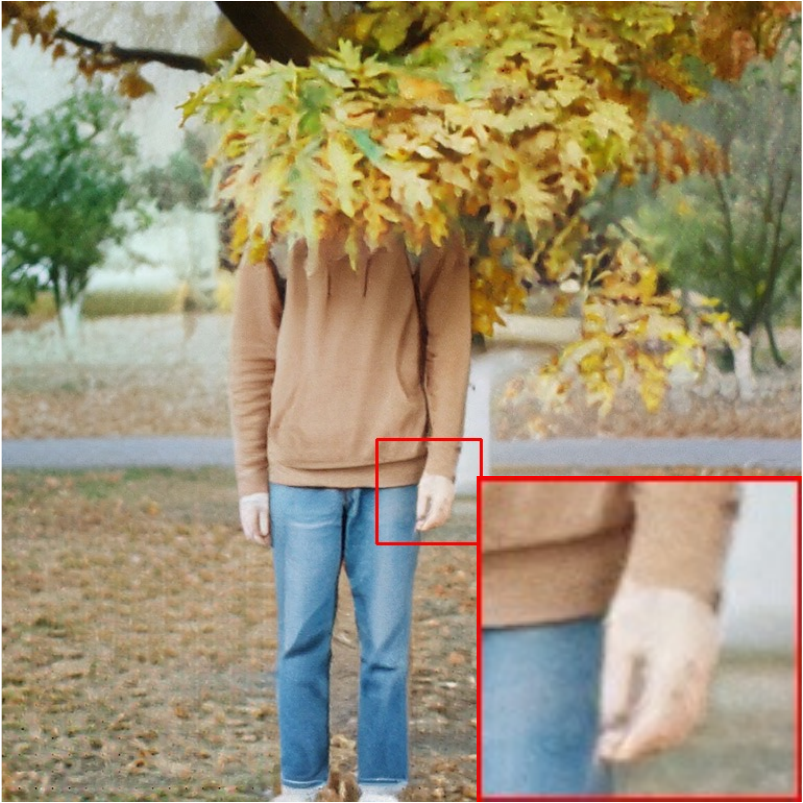}
\includegraphics[width=0.12\textwidth]{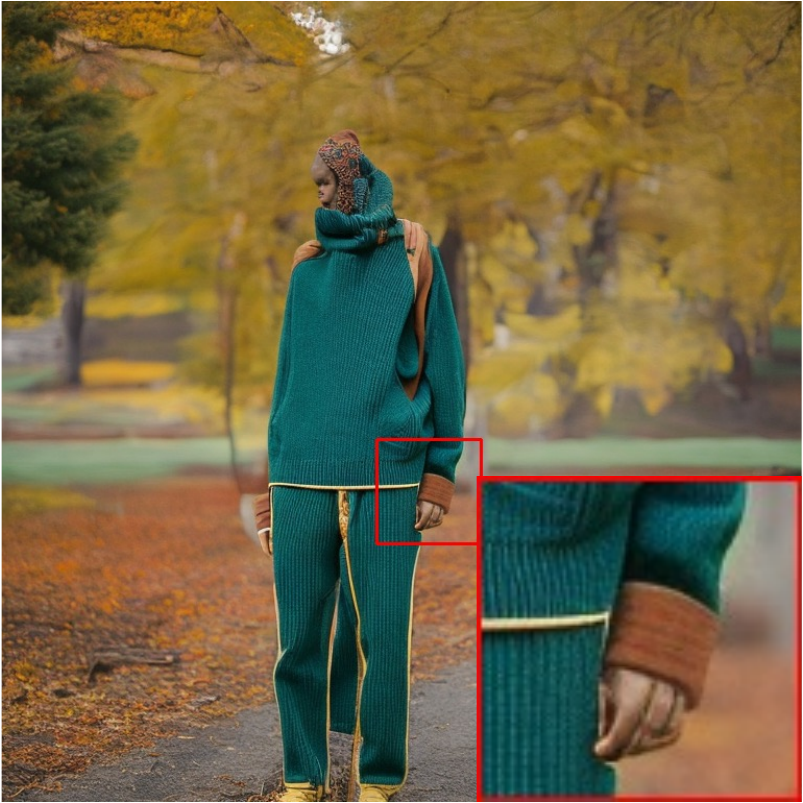}
\includegraphics[width=0.12\textwidth]{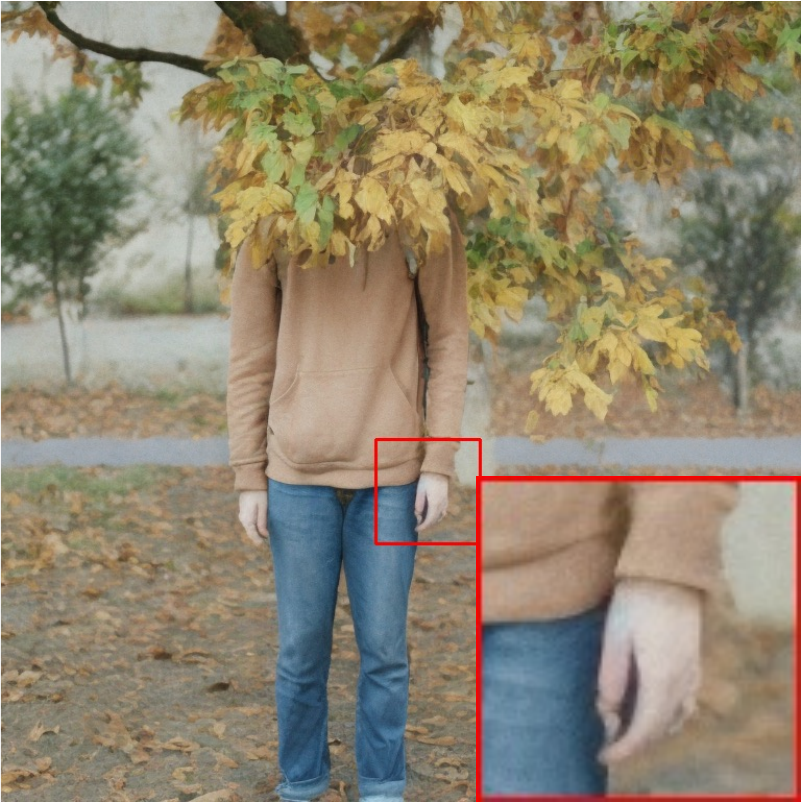}
\includegraphics[width=0.12\textwidth]{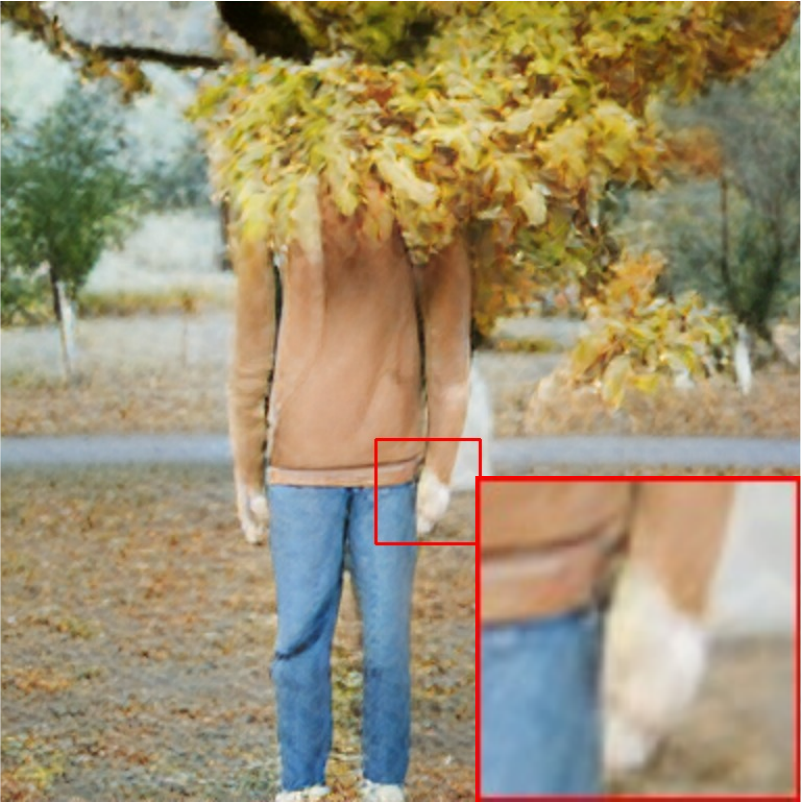} 
\includegraphics[width=0.12\textwidth]{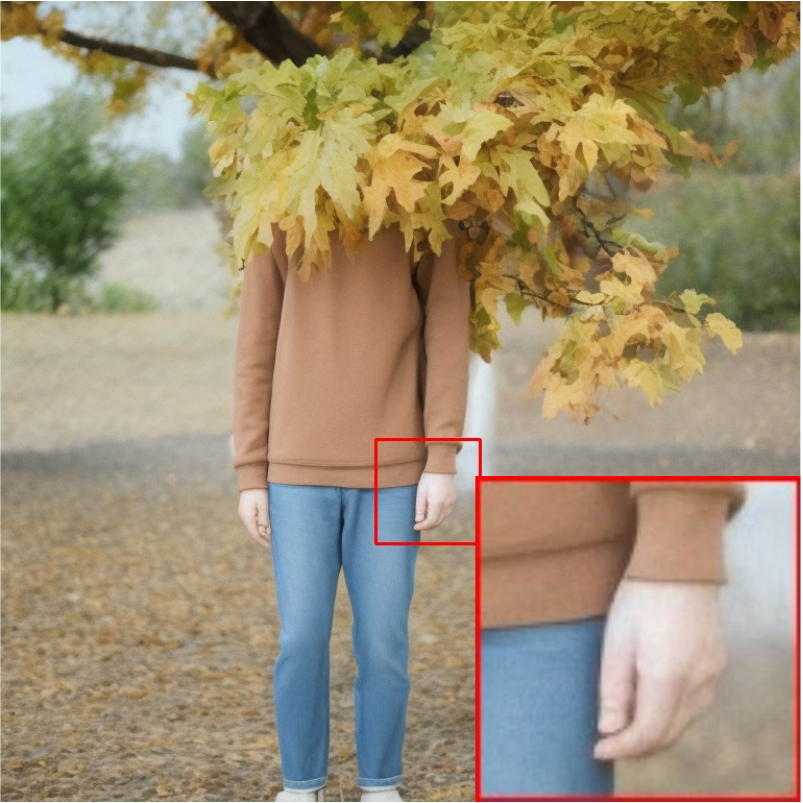} 
\\ \vspace{0.1cm}
\makebox[0.12\textwidth]{\textbf{bpp / DISTS$\downarrow$}}
\makebox[0.12\textwidth]{\textbf{0.0538 / 0.2610}}
\makebox[0.12\textwidth]{\textbf{0.0453 / 0.1789}}
\makebox[0.12\textwidth]{\textbf{0.0501 / 0.1207}}
\makebox[0.12\textwidth]{\textbf{0.0286 / 0.2688}}
\makebox[0.12\textwidth]{\textbf{0.0312 / 0.1651}}
\makebox[0.12\textwidth]{\textbf{0.0216 / 0.2046}}
\makebox[0.12\textwidth]{\textbf{0.0182 / 0.1804}}
\\
\caption{Visual comparisons of the proposed DiffEIC framework with the MSE-optimized ELIC \cite{ELIC}, the GANs-based HiFiC \cite{HiFiC} {and MS-ILLM \cite{MS-ILLM}}, the diffusion-based Text+Sketch \cite{Text+Sketch} and PerCo \cite{PerCo}, and the method by Wei et al. \cite{VQIR} on the CLIC2020 \cite{CLIC2020} dataset. For each method, the bpp and {DISTS} values are shown beneath images. Compared to other methods, our method produces more realistic and faithful reconstructions with lower bpp.}
\label{qualitative comparisons}
\end{figure*}

\subsection{Experimental Settings}
\subsubsection{Implementation}
We train DiffEIC on the \textbf{LSDIR} \cite{LSDIR} dataset, which contains 84,991 high-quality training images. The images are randomly cropped to 512$\times$512 resolution. In our experiments, we use Stable Diffusion 2.1-base\footnote{\url{https://huggingface.co/stabilityai/stable-diffusion-2-1-base}} as the diffusion prior. We train our model in an end-to-end manner using Eq. (\ref{loss}), where $\lambda_{sa}$ and $\lambda_{ne}$ are set to 2 and 1, respectively. To achieve different coding bitrates, we choose $\lambda$ from $\{1,2,4,8,16\}$. For optimization, we utilize Adam \cite{Adam} optimizer with $\beta_1=0.9$ and $\beta_2=0.999$ and set the learning rate to 1$\times 10^{-4}$. The training batch size is set to 4. Inspired by previous work \cite{Channel-wise}, we first train the proposed DiffEIC with $\lambda=1$ for 300K iterations, and then adapt them using target $\lambda$ for another 200K iterations. We set the learning rate to 2$\times 10^{-5}$ during the fine-tuning process. For inference, we adopt spaced DDPM sampling \cite{SpacedSampler} with 50 steps to reconstruct the images. All experiments are conducted on a single NVIDIA GeForce RTX 4090 GPU.

\subsubsection{Test Data}
For evaluation, we use three commonly used benchmarks: Kodak \cite{Kodak}, Tecnick \cite{Tecnick}, and CLIC2020 \cite{CLIC2020} datasets. The \textbf{Kodak} dataset contains 24 natural images with a resolution of 768$\times$512. The \textbf{Tecnick} dataset contains 140 images with 1200$\times$1200 resolution. The \textbf{CLIC2020} dataset has 428 high-quality images. {For the Tecnick and CLIC2020 datasets, we resize the images so that the shorter dimension is equal to 768px. Then we center-crop the image with 768$\times$768 resolution for evaluation~\cite{CDC}.}

\subsubsection{Metrics}
For quantitative evaluation, we employ several established metrics to assess the perceptual quality of results, including the Learned Perceptual Image Patch Similarity (\textbf{LPIPS}) \cite{LPIPS}, Naturalness Image Quality Evaluator (\textbf{NIQE}) \cite{NIQE}, {Deep Image Structure and Texture Similarity (\textbf{DISTS}) \cite{DISTS}}, Fréchet Inception Distance (\textbf{FID}) \cite{FID}, and Kernel Inception Distance (\textbf{KID}) \cite{KID}.
Meanwhile, we employ the Peak Signal-to-Noise Ratio (\textbf{PSNR}) and Multi-Scale Structural Similarity Index (\textbf{MS-SSIM}) \cite{MS-SSIM} to measure the fidelity of reconstruction results. Furthermore, the bits per pixel (bpp) is used to evaluate rate performance. Note that FID and KID are calculated on patches of 256$\times$256 resolution according to \cite{HiFiC}. Since the Kodak dataset is too small to calculate FID and KID, we do not report FID or KID results on it.

\subsection{Comparisons With State-of-the-art Methods}
We compare our DiffEIC with state-of-the-art learned image compression methods, including ELIC \cite{ELIC}, HiFiC \cite{HiFiC}, Text+Sketch \cite{Text+Sketch}, PerCo \cite{PerCo}, Wei et al. \cite{VQIR}, and MS-ILLM \cite{MS-ILLM}. For PerCo \cite{PerCo}, we use PerCo(SD)\footnote{\url{https://github.com/Nikolai10/PerCo/tree/master}} as a substitute, since the official source codes and models are not available. In addition, we compare with traditional image compression methods BPG \cite{BPG} and VVC \cite{VVC}. For BPG software, we optimize image quality and compression efficiency with the following settings: ``YUV444'' subsampling mode, ``x265'' HEVC implementation, ``8-bit'' depth, and ``YCbCr'' color space. For VVC, we employ the reference software VTM-23.0\footnote{\url{https://vcgit.hhi.fraunhofer.de/jvet/VVCSoftware_VTM/-/tree/master}} with intra configuration.

\begin{table*}[ht]
\renewcommand{\arraystretch}{1.3}
\centering
\scriptsize
\caption{Encoding and decoding speed on Kodak \cite{Kodak} dataset in terms of seconds.}
\label{comparition_time}
\begin{tabular}{ccc|
    S[table-format=5.3]@{\,\( \pm \)\,}S[table-format=1.3]
    S[table-format=5.3]@{\,\( \pm \)\,}S[table-format=1.3]
    c} 
\toprule
\textbf{Type} &\textbf{Method} &\textbf{Denoising Step}  &\multicolumn{2}{c}{\textbf{Encoding Speed (in sec.)}} &\multicolumn{2}{c}{\textbf{Decoding Speed (in sec.)}} &\textbf{Platform}\\
\midrule
Traditional method &VVC                 &--     & 13.862 & 9.821    & 0.066 & 0.006     &13th Core i9-13900K\\
\midrule
VAE-based method &ELIC                &--     & 0.056 & 0.006     & 0.081 & 0.011     &RTX4090\\
\midrule
\multirow{3}{*}{GAN-based methods} &HiFiC               &--     & 0.038 & 0.004     & 0.059 & 0.004     &RTX4090\\
 &MS-ILLM             &--     & 0.038 & 0.004     & 0.059 & 0.004     &RTX4090\\
 &Wei et al.          &--     & 0.050 & 0.003     & 0.179 & 0.005     &RTX4090\\
\midrule
\multirow{5}{*}{Diffusion-based methods} &Text+Sketch         &25     & 62.045 & 0.516    & 12.028 & 0.413    &RTX4090\\
 &PerCo               &5      & 0.080 & 0.018     & 0.665 & 0.009     &A100\\
 &PerCo               &20     & 0.080 & 0.018     & 2.551 & 0.018     &A100\\
 &DiffEIC (Ours)             &20     & 0.128 & 0.005     & 1.964 & 0.009     &RTX4090\\
 &DiffEIC (Ours)             &50     & 0.128 & 0.005     & 4.574 & 0.006     &RTX4090\\
\bottomrule
\end{tabular}
\end{table*}

\subsubsection{Quantitative Comparisons}
\label{Quantitative}
Fig. \ref{R-D-P} shows the rate-perception curves at low bitrates for different methods over the three datasets. 
It can be observed that the proposed DiffEIC performs much better than BPG \cite{BPG}, VVC \cite{VVC}, and ELIC \cite{ELIC} for all perceptual metrics.
Although the Text+Sketch \cite{Text+Sketch} achieves the best NIQE value of all the methods, it fails to ensure the pixel fidelity, where the LPIPS value is the highest.
For other generative image compression methods, the proposed DiffEIC yields lower DISTS, FID, and KID values, indicating that DiffEIC excels in preserving the perceptual integrity of the images and producing reconstructions with minimal perceptual differences from the originals.

The rate-distortion performance comparison is shown in Fig.~\ref{R-D}. Since Text+Sketch \cite{Text+Sketch} has ignored the pixel-level fidelity of the reconstruction results, we do not report its rate-distortion performance. Compared to Wei et al.~\cite{VQIR} and PerCo \cite{PerCo}, the proposed DiffEIC achieves better PSNR and MS-SSIM values. However, we find that DiffEIC is worse than other comparative methods. The reason behind this is that the proposed DiffEIC uses stable diffusion prior for realistic detail reconstruction at extremely low bitrates, which does not ensure pixel-level accuracy. To further demonstrate this, we report the PSNR and MS-SSIM values of the stable diffusion autoencoder (see the black horizontal line in Fig.~\ref{R-D}), which can be treated as the upper bound of the performance of DiffEIC. Although it sacrifices some fidelity, the proposed DiffEIC effectively capture realism at extremely low bitrates. 

\subsubsection{Qualitative Comparisons}
Fig. \ref{qualitative comparisons} shows visual comparisons among the evaluated methods at extremely low bitrates. 
Compared to other methods, DiffEIC yields reconstructions with higher perceptual quality, fewer artifacts, and more realistic detail at extremely low bitrates. For example, DiffEIC preserves the texture and details of the background that are lost or distorted in other methods (see the first row). Similarly, the DiffEIC is able to produce more realistic facial detail than other methods (see the second row).

\subsubsection{{Complexity Comparisons}}
We further compare the proposed DiffEIC with state-of-the-art image compression methods in terms of complexity. For PerCo~\cite{PerCo}, we directly show the results reported in their paper, since the official source codes are not available. Table~\ref{comparition_time} summarizes the average encoding/decoding time in seconds with its standard deviation on the Kodak dataset. On the one hand, it is worth noting that the diffusion-based methods have higher encoding and decoding complexity than the VAE-based and GAN-based methods. On the other hand, the proposed DiffEIC encoder is significantly faster than Text+Sketch \cite{Text+Sketch}. Compared to PerCo~\cite{PerCo}, the proposed DiffEIC is able to achieve comparable encoding speed and faster decoding speed with the same number of denoising steps.

%% file: 06_analysis.tex
\section{Analysis and Discussions}
\label{analysis}
To better analyze the proposed method, we perform ablation studies and discuss its limitations.

\subsection{Ablation of Latent Feature Guidance}
\label{LFG_effectiveness}
In this part, we analyze the proposed Latent Feature Guidance (LFG), which is used to correct content variables. Specifically, we remove the guidance components and retrain the model from scratch using the same experimental settings.

\begin{figure}[thbp]\scriptsize
\centering
\includegraphics[width=0.24\textwidth]{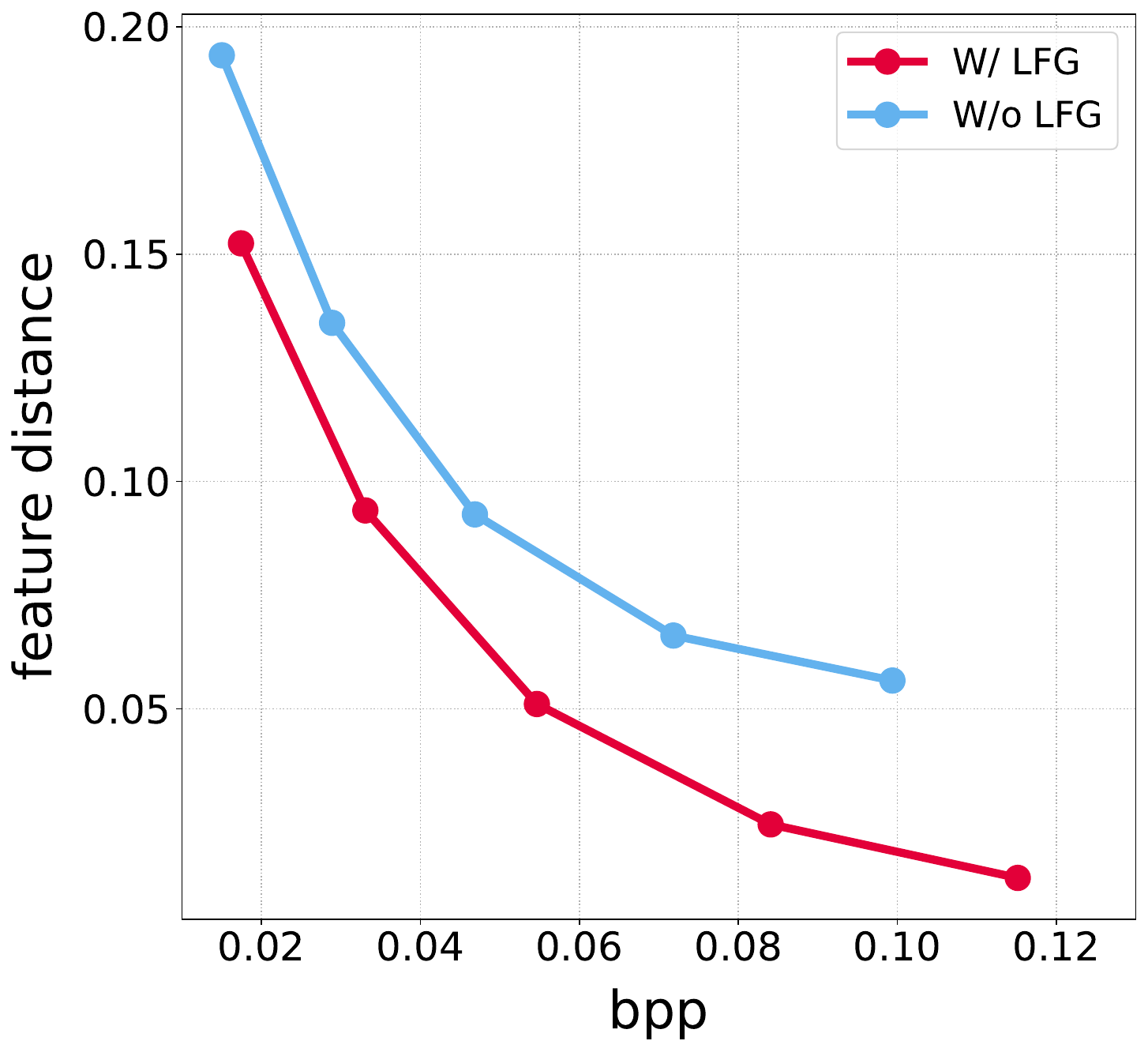}
\includegraphics[width=0.24\textwidth]{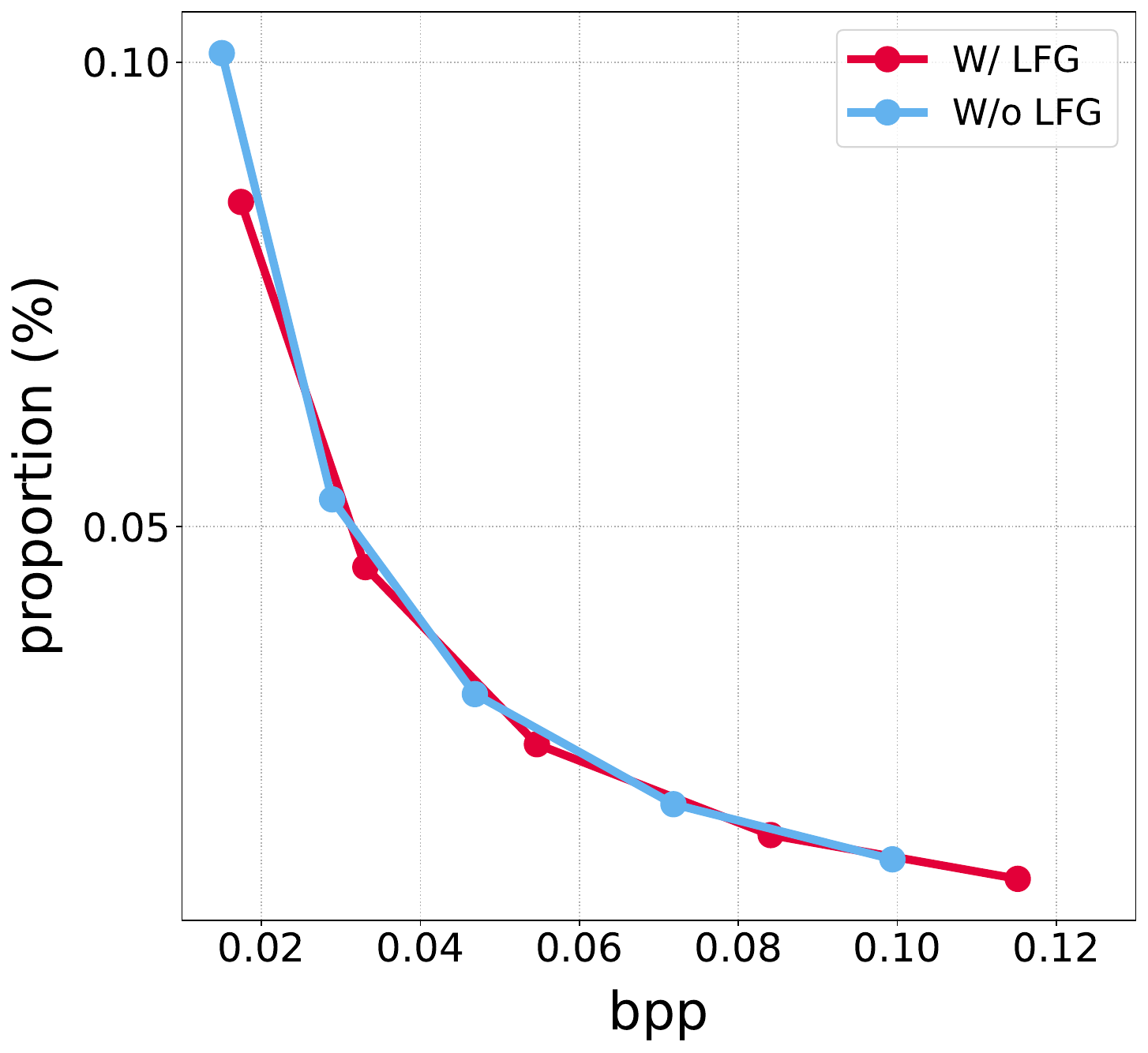} \\ \vspace{0.1cm}
\makebox[0.01\textwidth]{}
\makebox[0.23\textwidth]{\textbf{(a)}}
\makebox[0.005\textwidth]{}
\makebox[0.23\textwidth]{\textbf{(b)}}\\
\caption{Ablation studies of latent feature guidance on CLIC2020 \cite{CLIC2020} dataset. (a) Euclidean distance between content variables and corresponding latent representations; {(b) Proportion of bits allocated to the hyper prior.}} 
\label{FD}
\end{figure}

\begin{figure}[thbp]\scriptsize
\centering
\makebox[0.16\textwidth]{(a) \textbf{Original}}
\makebox[0.16\textwidth]{(b) \textbf{W/o LFG}}
\makebox[0.16\textwidth]{(c) \textbf{Ours}}\\ \vspace{0.1cm}
\includegraphics[width=0.16\textwidth]{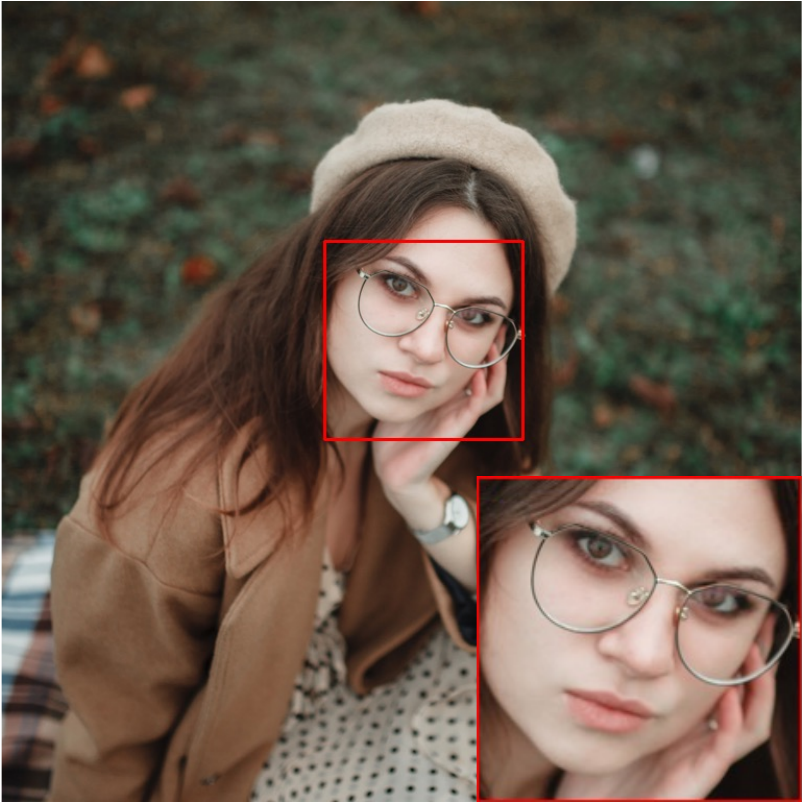}
\includegraphics[width=0.16\textwidth]{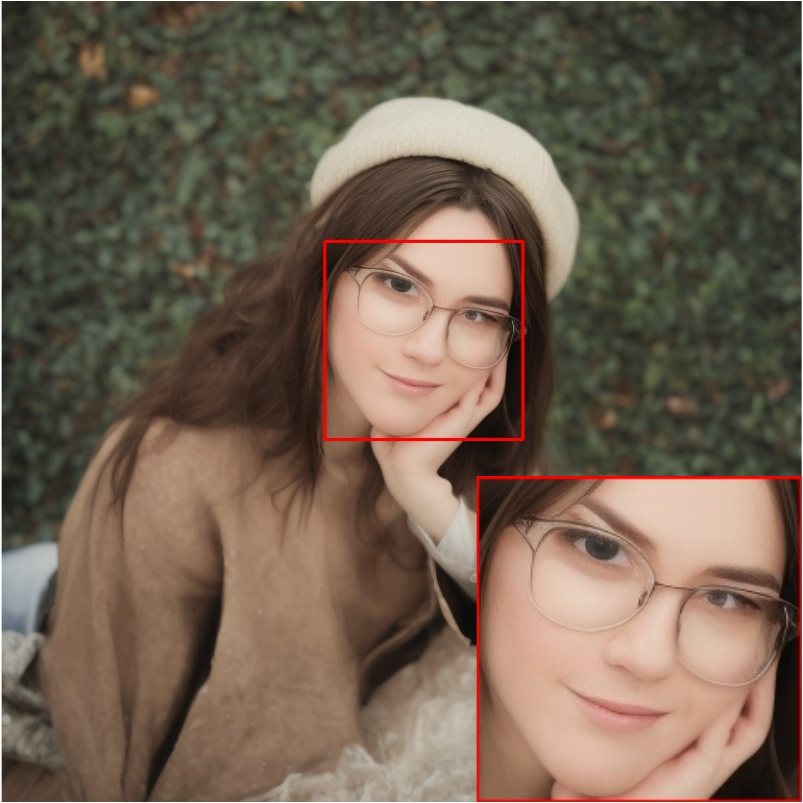}
\includegraphics[width=0.16\textwidth]{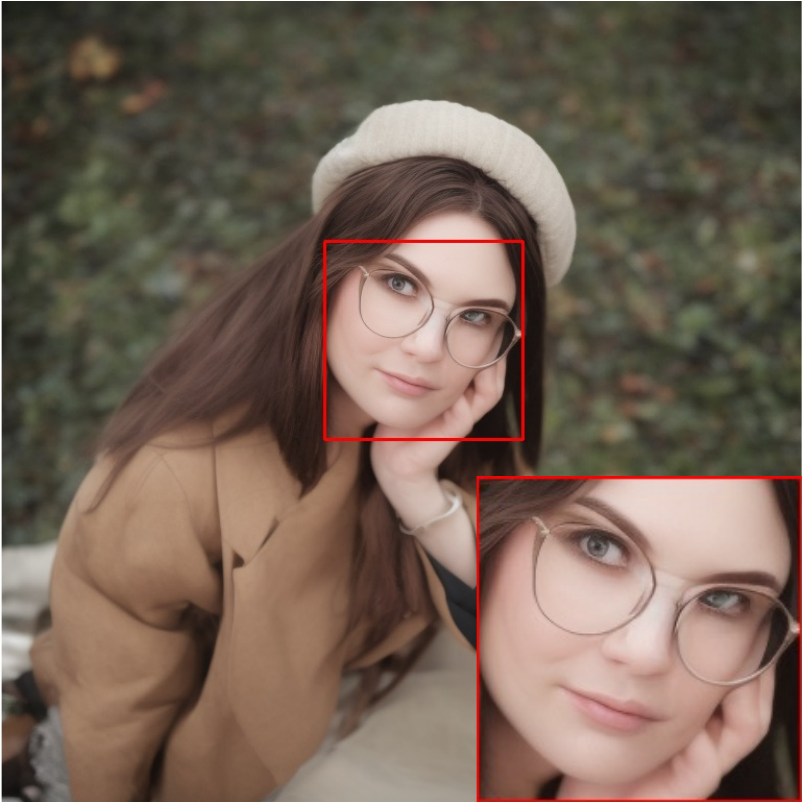}
\\ \vspace{0.1cm}
\makebox[0.16\textwidth]{\textbf{bpp / MS-SSIM$\uparrow$ / DISTS$\downarrow$}}
\makebox[0.16\textwidth]{\textbf{0.0076 / 0.55 / 0.2231}}
\makebox[0.16\textwidth]{\textbf{0.0085 / 0.60 / 0.2000}}\\
\caption{Impact of latent feature guidance on reconstruction results.} 
\label{LFG}
\end{figure}

\begin{table*}[ht]
\renewcommand{\arraystretch}{1.3}
\centering
\scriptsize
\caption{Ablation of Latent Feature Guidance (LFG), Denoising Steps (DS), and the Channel Number (CN) of the control module. BD-rate \cite{BD-rate} is calculated on CLIC2020 \cite{CLIC2020} dataset, with DISTS and LPIPS as the metric.}
\label{table_ablation}
\begin{tabular}{lcc|cc|cc}
\toprule
\multirow{2}{*}{\textbf{Methods}} & \multicolumn{2}{c}{\textbf{Hyper-Parameter}} &\multicolumn{2}{c}{\textbf{BD-Rate (\%)}} &\multicolumn{2}{c}{\textbf{Speed (in sec.)}} \\
\cline{2-7}
& CN (\%) &Denoising Steps  &DISTS &LPIPS  &Encoding Speed &Decoding Speed\\
\midrule
\multirow{4}{*}{DiffEIC (CN)}    
                    &100    &50     &-5.36   &-2.57       &\multirow{4}{*}{0.128 $\pm$ 0.005}     & 6.012 $\pm$ 0.012\\
                    &50     &50     &-2.99  &-2.28  &     & 5.068 $\pm$ 0.020\\
                    &100    &37     &4.11   &0.80     &     & 4.578 $\pm$ 0.013\\
                    &50     &43     &0.96  &0.19  &     & 4.474 $\pm$ 0.019\\
\midrule
DiffEIC (W/o LFG)   &20     &50     &23.88  &13.19  & 0.062 $\pm$ 0.009     & 4.574 $\pm$ 0.006\\
DiffEIC (Ours)      &20     &50     &0      &0      & 0.128 $\pm$ 0.005     & 4.574 $\pm$ 0.006\\
\midrule
\multirow{4}{*}{DiffEIC (DS)} 
                    &20     &20     &22.51  &6.20  &\multirow{4}{*}{0.128 $\pm$ 0.005}     & 1.964 $\pm$ 0.009\\
                    &20     &10     &37.83  &13.26  &     & 1.089 $\pm$ 0.009\\
                    &20     &5      &49.93  &21.68  &     & 0.646 $\pm$ 0.005\\
                    &20     &0      &59.50  &35.77  &     & 0.212 $\pm$ 0.006\\
\bottomrule
\end{tabular}
\end{table*}

Fig. \ref{FD}(a) demonstrates that the distance between the content variables and corresponding latent representations is significantly reduced after introducing the LFG strategy, which implies that more accurate information is provided for the subsequent denoising process. As shown in Table \ref{table_ablation}, the removal of guidance components results in a slightly faster encoding speed but a noticeable degradation in performance, with a 23.88\% increase in bitrates at the same DISTS metric and a 13.19\% increase in bitrates at the same LPIPS metric. The visual comparison is presented in Fig. \ref{LFG}. As seen from this example, with the help of LFG strategy, our DiffEIC achieves more accurate facial reconstruction at extremely low bitrates. This further demonstrates that the LFG strategy contributes to increased fidelity.

Note that the representation $w$ in the decoder side is extracted from the quantized side information $\hat{z}$. The bitrates of the additional information contained in $\hat{z}$ need to be analyzed further. To evaluate the impact of the LFG on the bitrates of the hyper prior, we compare the bits allocated to the hyper prior with and without LFG. Specifically, we compute the proportion of bits allocated to hyper prior as:
\begin{equation}
\label{proportion_calculation} 
    P = \frac{R(\hat{z})}{R(\hat{z})+R(\hat{y})}
\end{equation}
where $\mathcal{R}(\cdot)$ is the bitrate. As shown in Fig. \ref{FD}(b), using LFG does not significantly affect the proportion of bits allocated to the hyper prior. The reason for this phenomenon is that the hyper prior $\hat{z}$ requires extremely fewer bits than the quantized latent representation $\hat{y}$, which is also observed in~\cite{Hyperprior}, so the additional information conveyed by the hyper prior is small and the bit consumption of these information can be ignored. 

We further explore different fusion methods in the proposed LFGCM, including the addition, concatenation, and cross-attention. 
As shown in Table \ref{fusion}, the fusion method using addition causes a severe degradation in  compression performance. In addition, using the cross-attention mechanism achieves better compression performance than our method, while it increases the computational complexity by 31\% in terms of MACs. 
Considering the trade-off between performance and computational complexity, 
we choose the concatenation method for feature fusion in the proposed DiffEIC.

\begin{table}[ht]
\renewcommand{\arraystretch}{1.3}
\centering
\scriptsize
\caption{Ablation studies of the fusion method in LFGCM. BD-Rate is evaluated on CLIC2020 dataset and the MACs of different fusion methods are calculated based on 768$\times$512 image patch.}
\label{fusion}
\begin{tabular}{lccccc}
\toprule
\multirow{2}{*}{\textbf{Methods}} &\multicolumn{4}{c}{\textbf{BD-Rate (\%)}} &\multirow{2}{*}{\textbf{MACs (G)}}\\
\cline{2-5}
&DISTS &LPIPS &NIQE &MS-SSIM \\
\midrule
Addition &6.66 &5.24 &20.46 &2.43 &1.13 (-17\%)\\
Cross-Attention &-0.45 &-0.67 &-2.08 &-1.58 &1.79 (+31\%)\\
Concatenation (Ours) & 0 & 0 & 0 & 0 &1.36 (+ 0\%)\\
\bottomrule
\end{tabular}
\end{table}

\begin{figure*}[htbp]\scriptsize
\centering
\includegraphics[width=0.85\textwidth]{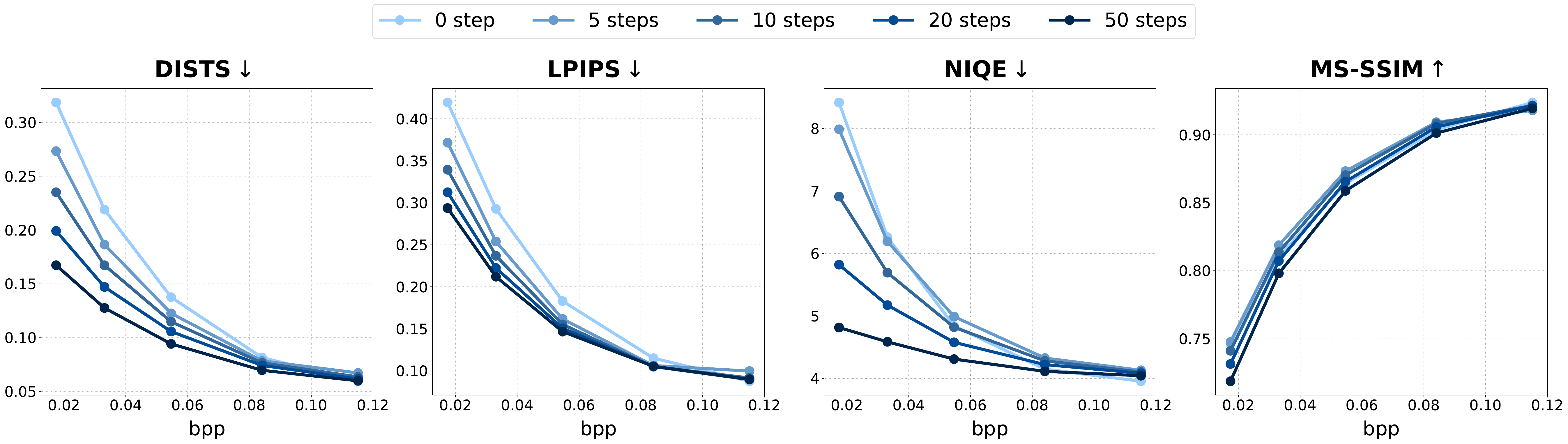} 
\caption{{Quantitative comparisons of different denoising steps on the CLIC2020 \cite{CLIC2020} dataset. \textit{0 step} denotes directly using the decoder $\mathcal{D}$ to decode content variables.}}
\label{QC_of_step}
\end{figure*}

\begin{figure*}[htbp]\scriptsize
\centering
\makebox[0.015\textwidth]{}
\makebox[0.137\textwidth]{(a) \textbf{Original}}
\makebox[0.137\textwidth]{(b) \textbf{0 step}}
\makebox[0.137\textwidth]{(c) \textbf{5 steps}}
\makebox[0.137\textwidth]{(d) \textbf{10 steps}}
\makebox[0.137\textwidth]{(e) \textbf{20 steps}}
\makebox[0.137\textwidth]{(f) \textbf{50 steps}}
 \\ \vspace{0.1cm}
 \parbox[b][0.137\textwidth][c]{0.015\textwidth}{\rotatebox{90}{\textbf{0.0739 bpp}}}
\includegraphics[width=0.137\textwidth]{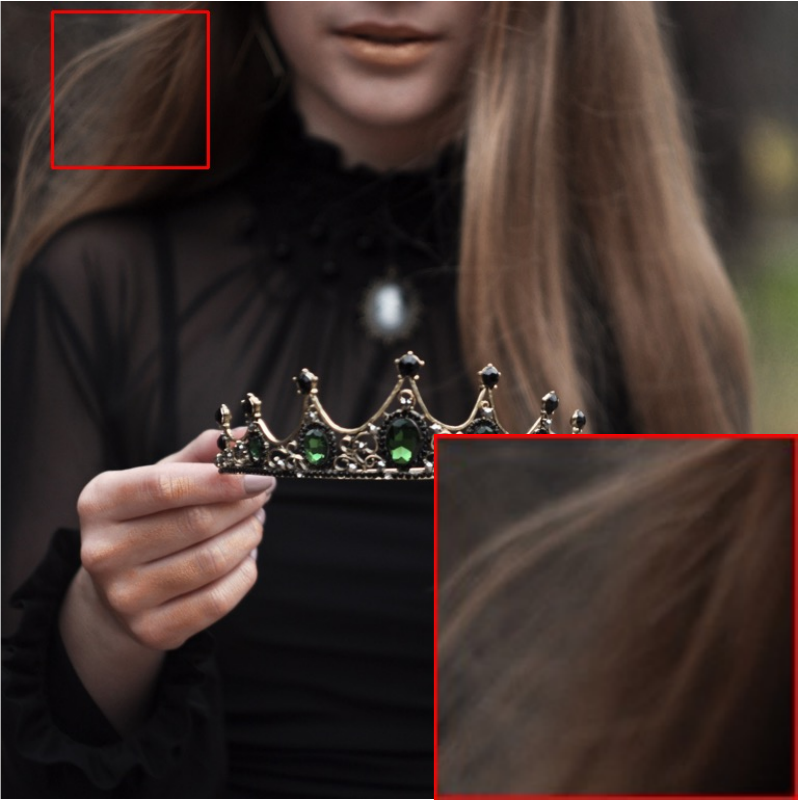}
\includegraphics[width=0.137\textwidth]{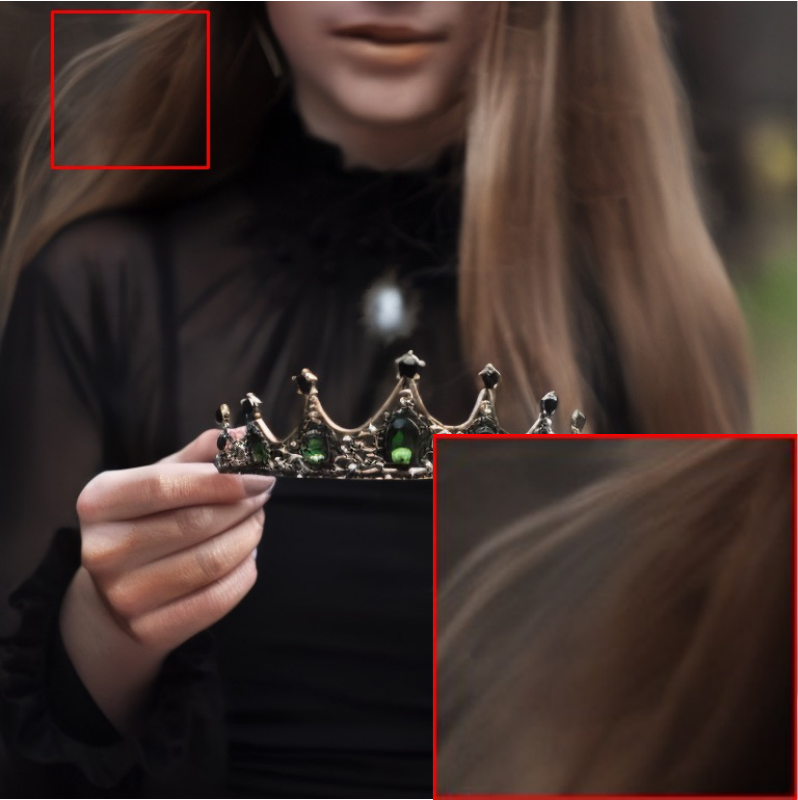}
\includegraphics[width=0.137\textwidth]{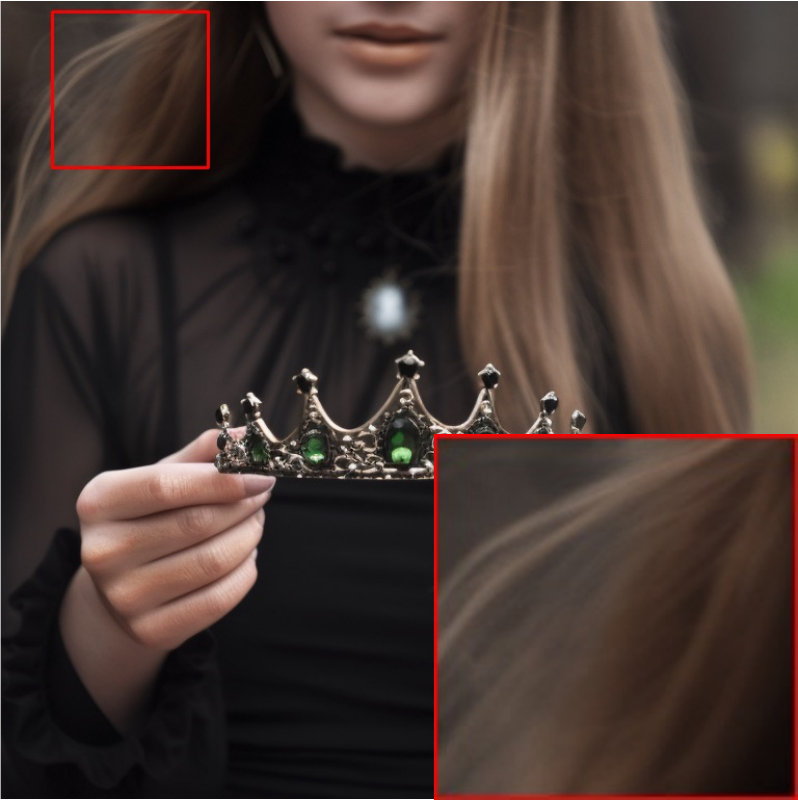}
\includegraphics[width=0.137\textwidth]{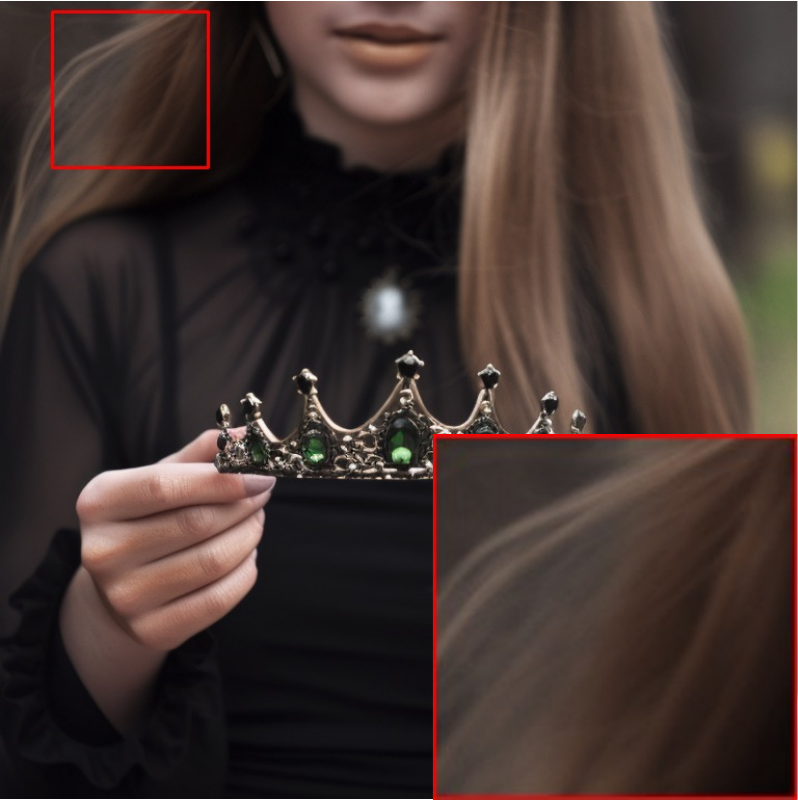}
\includegraphics[width=0.137\textwidth]{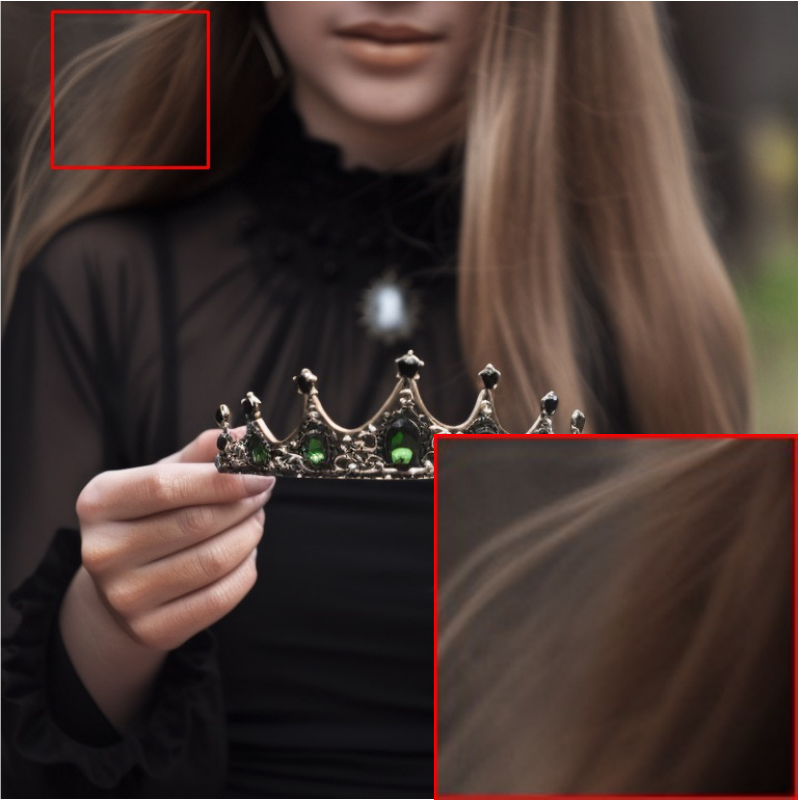}
\includegraphics[width=0.137\textwidth]{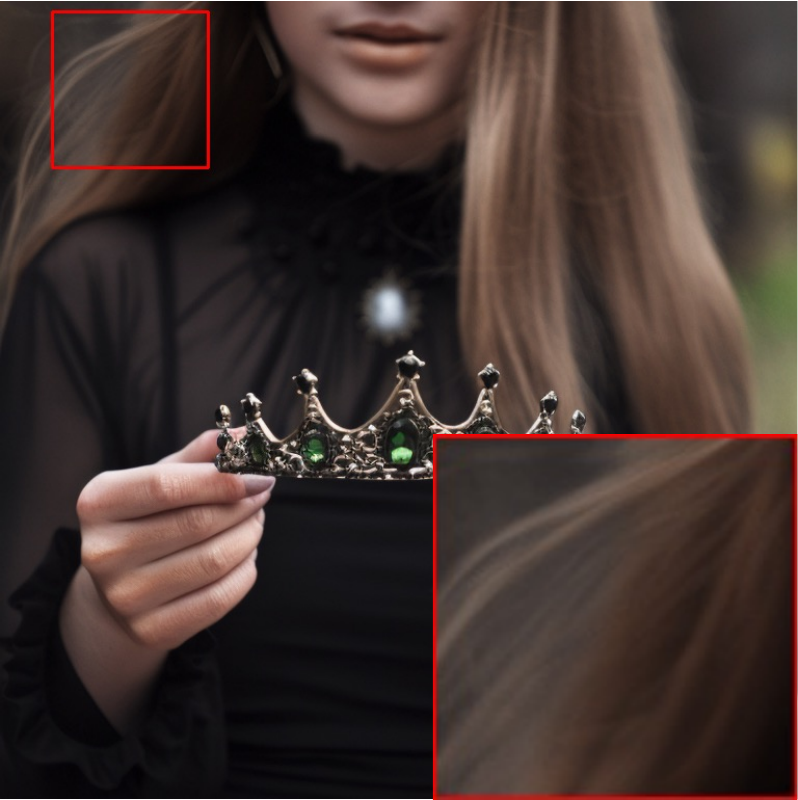} 
\\ \vspace{0.1cm}
\makebox[0.015\textwidth]{}
\makebox[0.137\textwidth]{\textbf{MS-SSIM$\uparrow$ / DISTS$\downarrow$}}
\makebox[0.137\textwidth]{\textbf{0.9682 / 0.0824}}
\makebox[0.137\textwidth]{\textbf{0.9695 / 0.0748}}
\makebox[0.137\textwidth]{\textbf{0.9703 / 0.0728}}
\makebox[0.137\textwidth]{\textbf{0.9667 / 0.0700}}
\makebox[0.137\textwidth]{\textbf{0.9662 / 0.0677}}
\\ \vspace{0.1cm}
\parbox[b][0.137\textwidth][c]{0.015\textwidth}{\rotatebox{90}{\textbf{0.0174 bpp}}}
\includegraphics[width=0.137\textwidth]{figs/sample_step/original.pdf}
\includegraphics[width=0.137\textwidth]{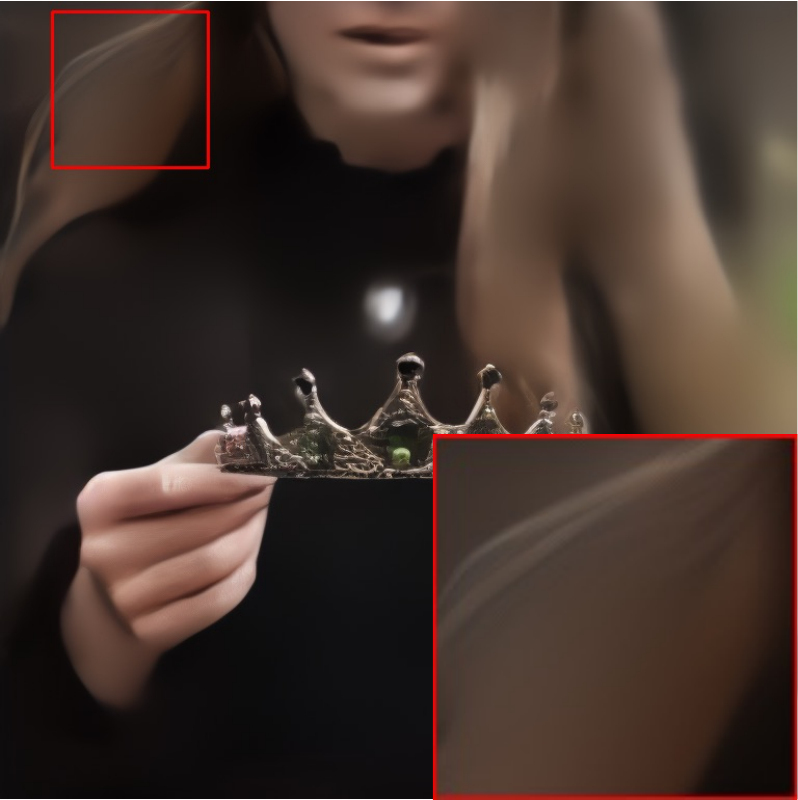}
\includegraphics[width=0.137\textwidth]{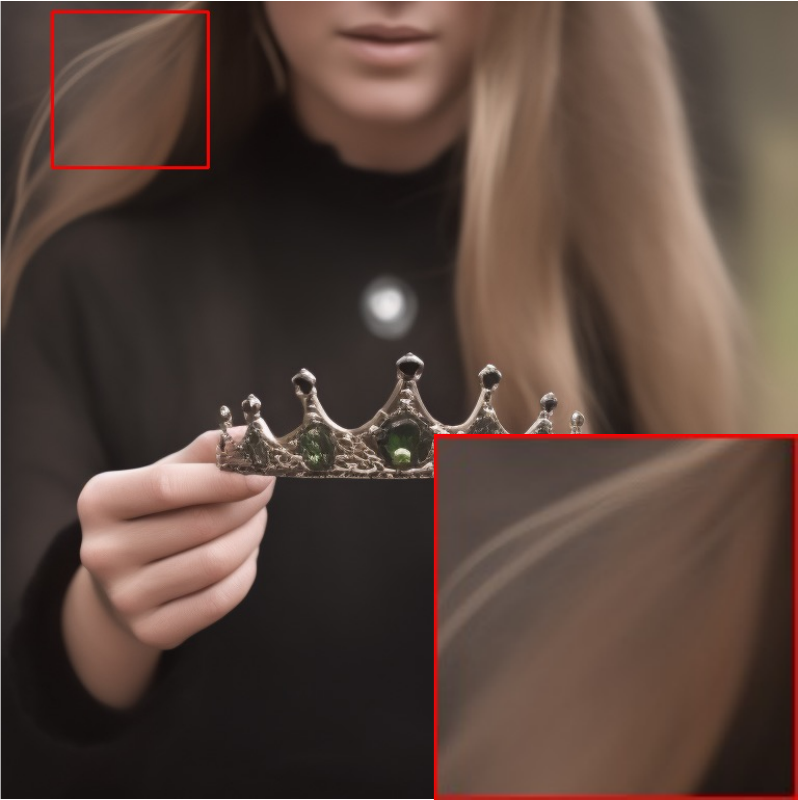}
\includegraphics[width=0.137\textwidth]{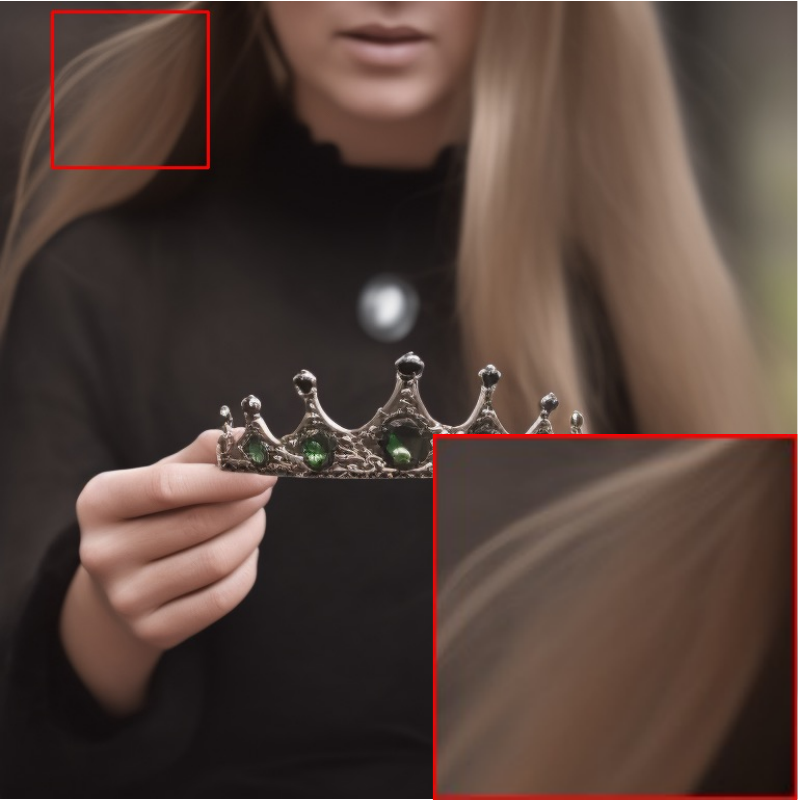}
\includegraphics[width=0.137\textwidth]{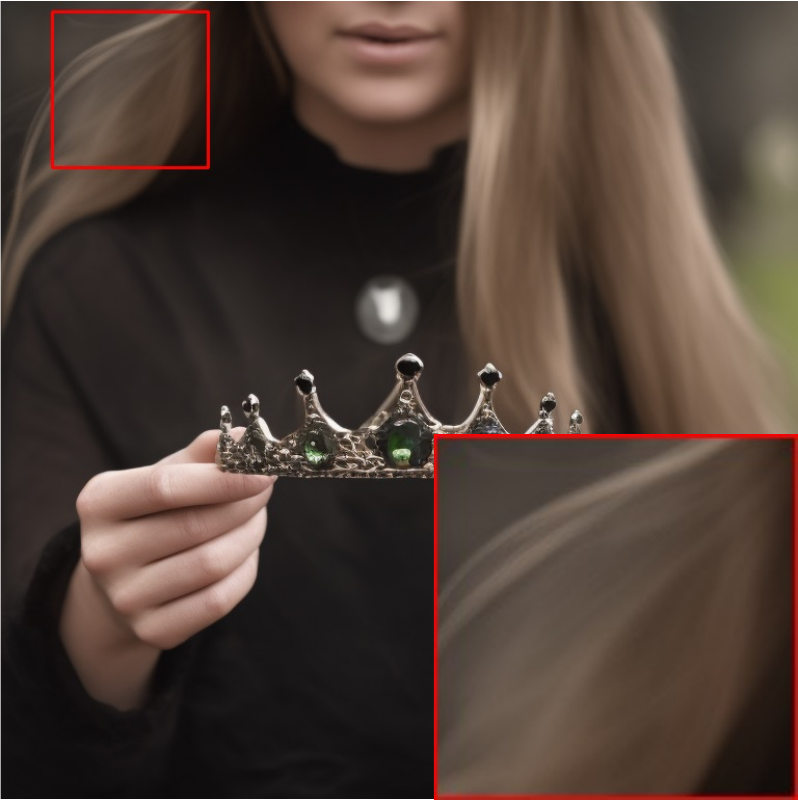}
\includegraphics[width=0.137\textwidth]{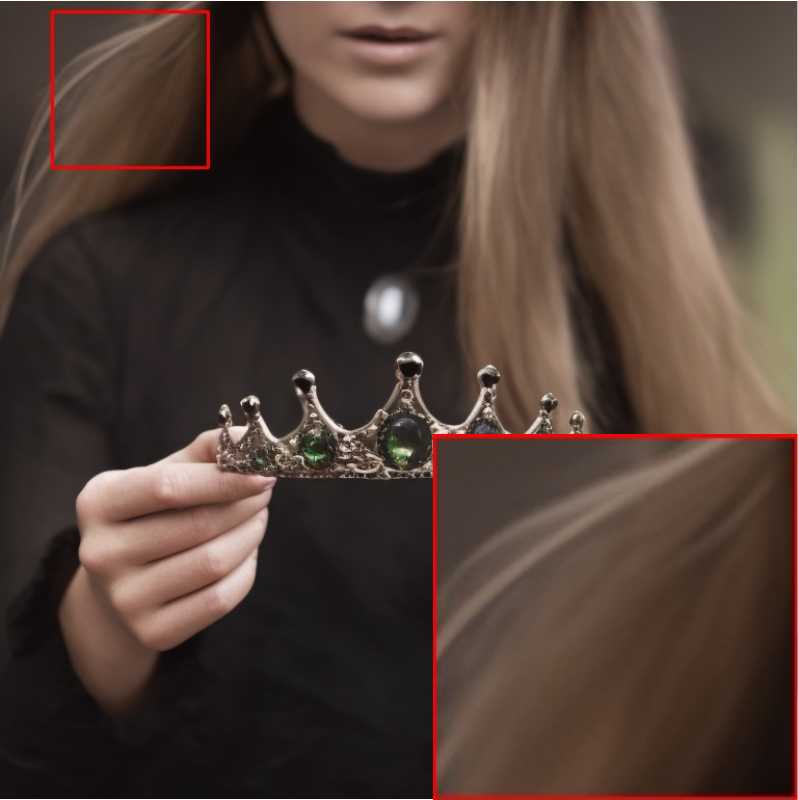} 
\\ \vspace{0.1cm}
\makebox[0.015\textwidth]{}
\makebox[0.137\textwidth]{\textbf{MS-SSIM$\uparrow$ / DISTS$\downarrow$}}
\makebox[0.137\textwidth]{\textbf{0.8842 / 0.2585}}
\makebox[0.137\textwidth]{\textbf{0.8733 / 0.2293}}
\makebox[0.137\textwidth]{\textbf{0.8755 / 0.1993}}
\makebox[0.137\textwidth]{\textbf{0.8649 / 0.1752}}
\makebox[0.137\textwidth]{\textbf{0.8641 / 0.1701}}
\caption{{Visual comparisons of different denoising steps.}}
\label{VC_of_step}
\end{figure*}

\subsection{{Effect of the Channel Number in Control Module}}
\label{ablation_of_cn}
We further analyze how the number of channels of the control module affects the performance and complexity of the proposed DiffEIC. In our default setting, we reduce the number of channels to 20\% of the original. We also increase the number of channels by setting the percentage to 50\% and 100\%. As shown in Table~\ref{table_ablation}, using more channels is able to bring a slight improvement in performance, where the lower DISTS and LPIPS values are achieved. However, it inevitably leads to the decoding complexity up. For example, the decoding speed of the proposed DiffEIC with 100\% channels decrease by about 31\% compared to the default setting. When the decoding time is comparable, using 20\% channels results in slightly better performance than using higher percentages of channels. This indicates that reducing the number of channels is more effective in balancing performance and inference speed than simply reducing the number of denoising steps. To achieve a tradeoff between performance and inference speed, we choose the 20\% in the proposed DiffEIC.

\subsection{{Effect of Denoising Steps}}
\label{denoising_step}
For the proposed DiffEIC, we relate the decoding complexity to the number of denoising steps. As shown in Table~\ref{table_ablation}, the decoding complexity can be reduced by using fewer denoising steps. Fig.~\ref{QC_of_step} shows the reconstruction performance using different numbers of denoising steps. We note that increasing the number of denoising steps is able to improve the perceptual quality of the decoded results, where the perceptual metrics (DISTS, LPIPS, and NIQE) are better. The visual comparisons in Fig.~\ref{VC_of_step} further demonstrate that using more denoising steps facilitates the improvement of the reconstruction performance, where the details of the hair are well reconstructed.

\subsection{Effectiveness of Space Alignment Loss}
\label{analysis_sal}
{The proposed space alignment loss is used to provide constraints for LFGCM. To illustrate the necessity of this loss, we attempt to train DiffEIC without the space alignment loss by removing $\mathcal{L}_{sa}$ from Eq. (\ref{loss}).}

\begin{figure}[tbp]\scriptsize
\centering
\includegraphics[width=0.24\textwidth]{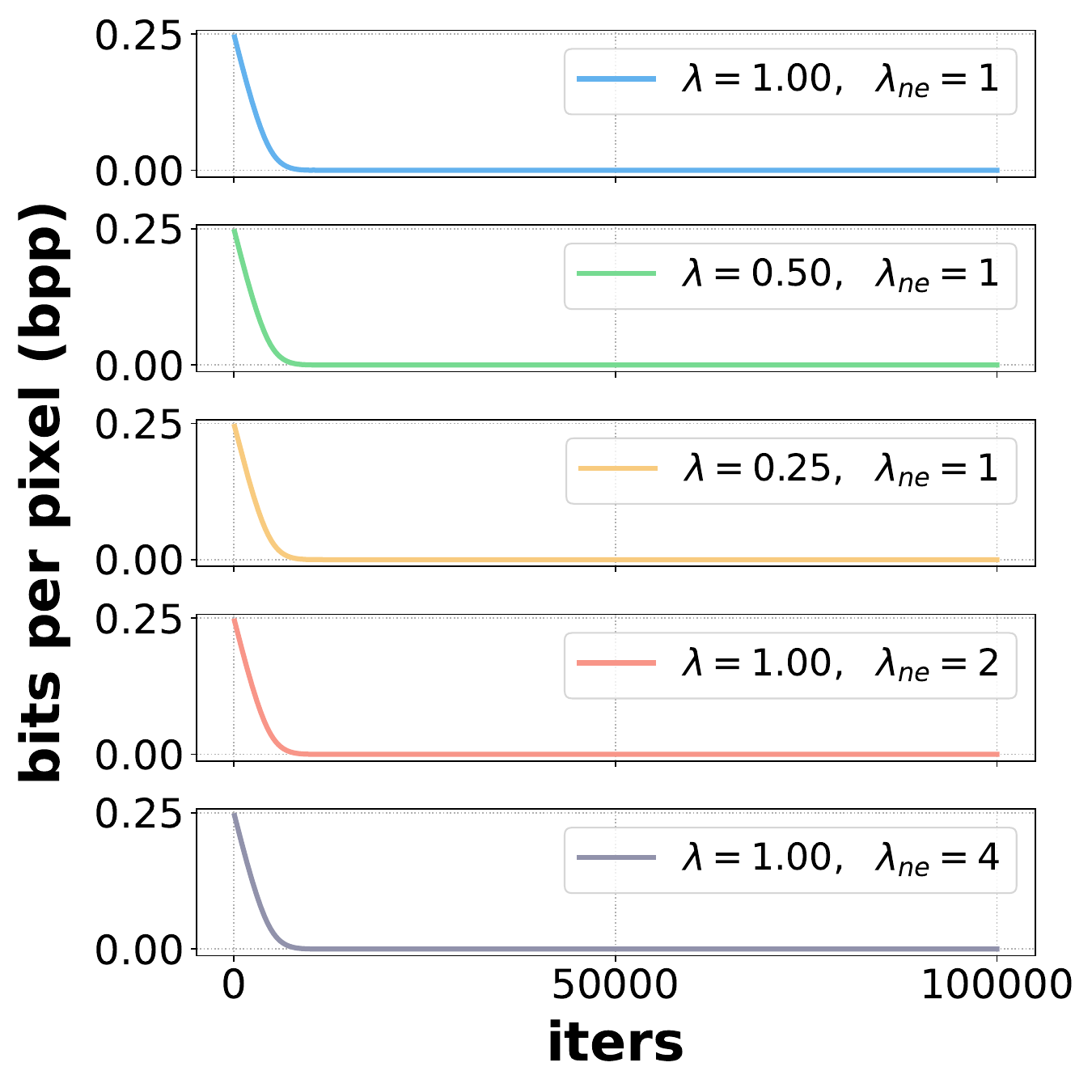}
\includegraphics[width=0.24\textwidth]{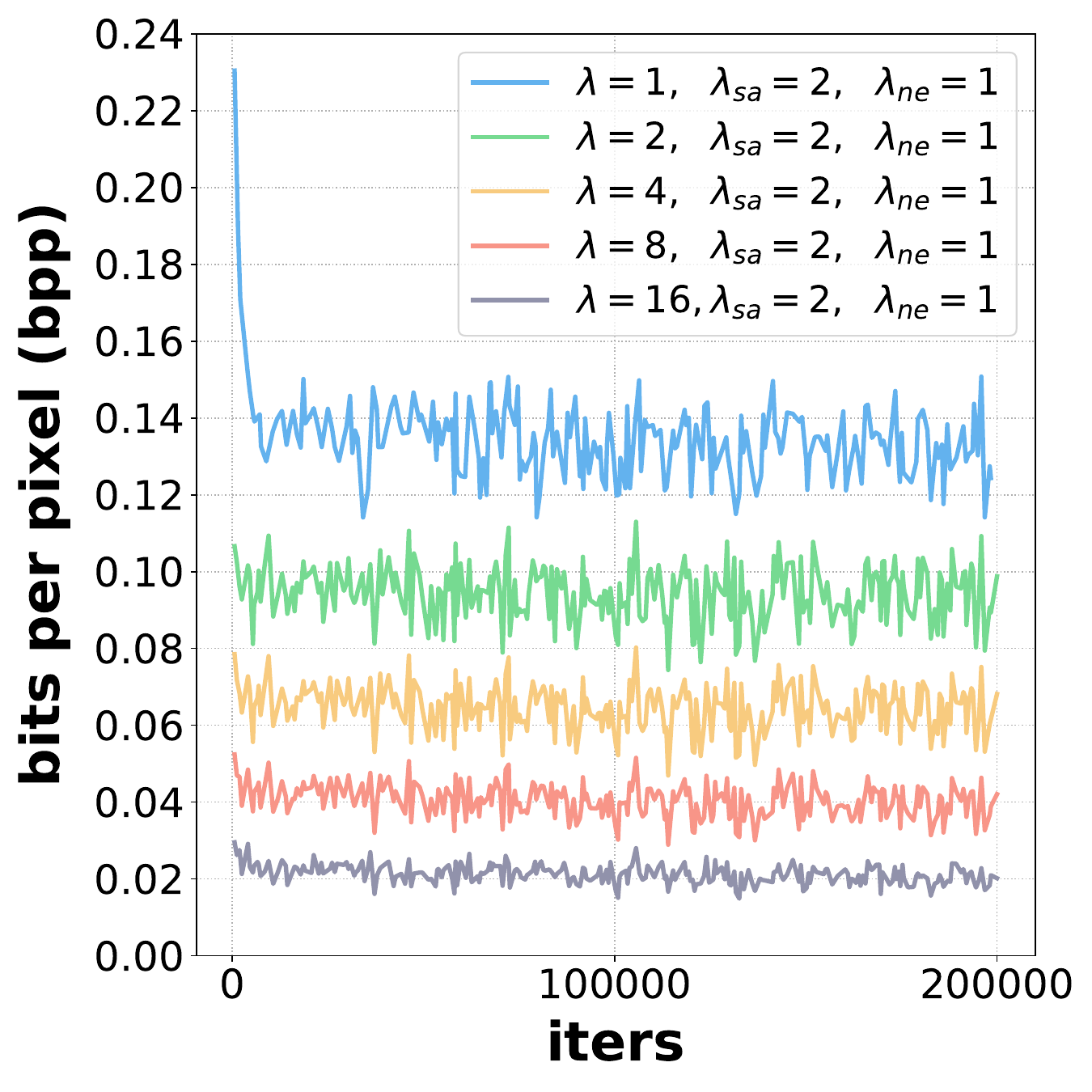}
\vspace{0.1cm}\\
\makebox[0.24\textwidth]{(a) W/o space alignment loss}
\makebox[0.24\textwidth]{(b) W/ space alignment loss}\\
\caption{Effectiveness of the space alignment loss for end-to-end training.} 
\label{SAL}
\end{figure}

As shown in Fig. \ref{SAL}(a), {without the space alignment loss,} the bits per pixel (bpp) curves invariably converge to zero during training, regardless of the selected values for $\lambda$ and $\lambda_{ne}$. We attribute this phenomenon to the noise estimation loss being {independent of the input images, thus failing} to provide effective constraints for LFGCM.
In contrast, Fig. \ref{SAL}(b) demonstrates the effectiveness of incorporating the space alignment loss. With this loss in place, the bpp curves stabilize and converge to meaningful values during training, indicating that the space alignment loss successfully enforces necessary constraints. Furthermore, the space alignment loss forces content variables to align with the diffusion space, contributing to enhanced reconstruction quality, as mentioned in Section \ref{LFG_effectiveness}.

\subsection{{Robustness to Different Image Resolutions}}
\begin{figure}[htbp]\scriptsize
\centering
\includegraphics[width=0.24\textwidth]{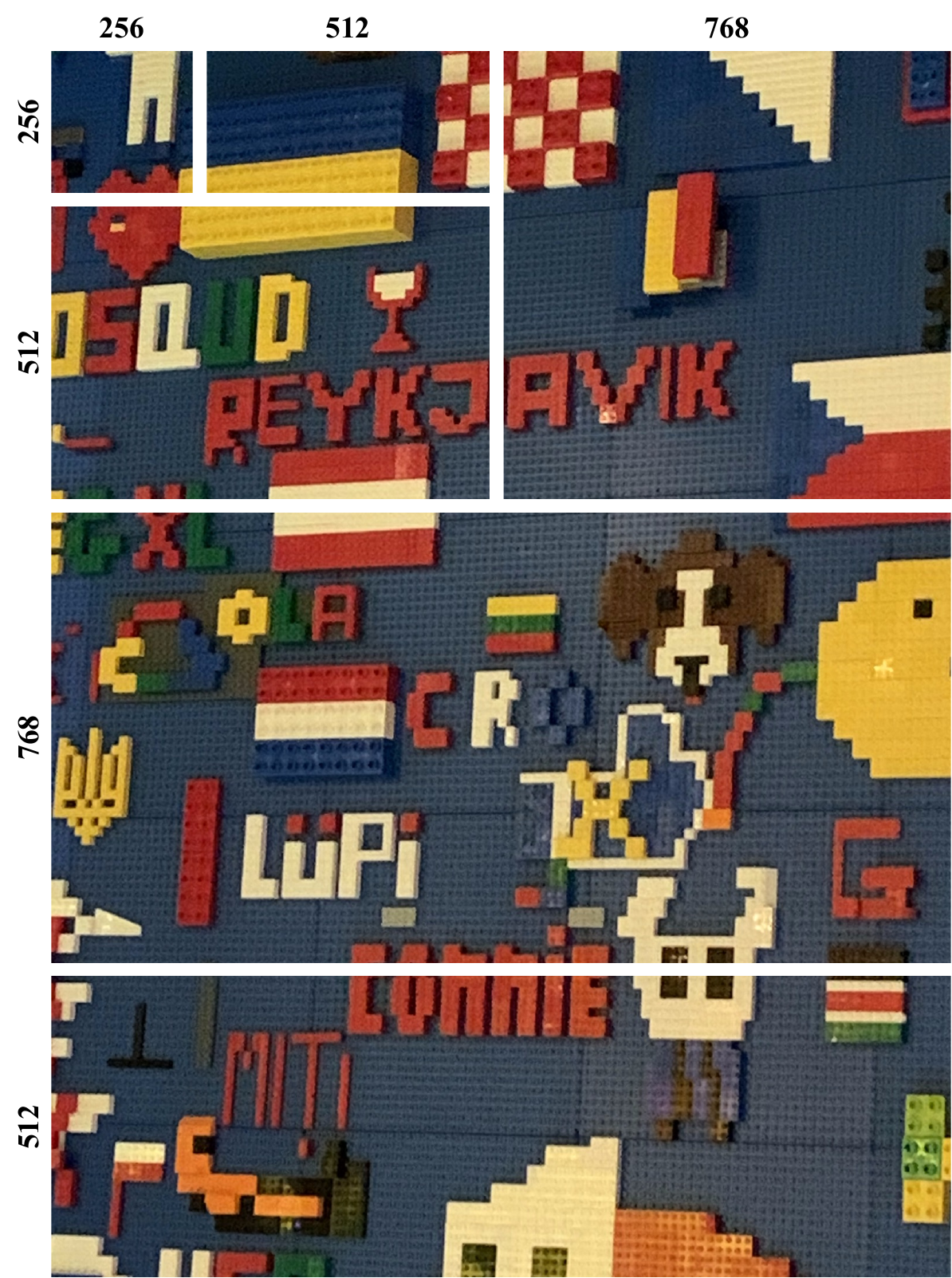}
\includegraphics[width=0.24\textwidth]{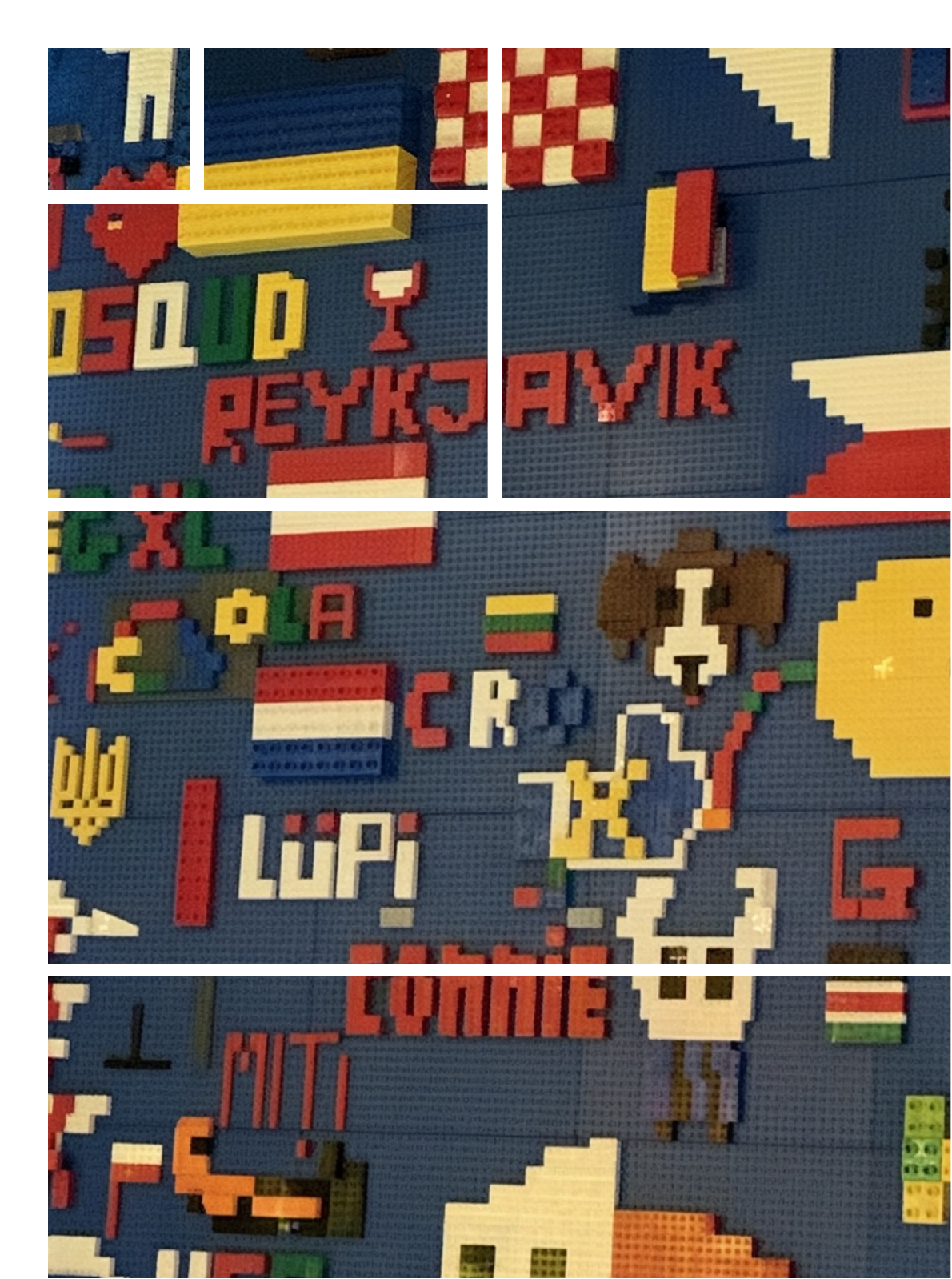}
\\ \vspace{0.1cm}
\makebox[0.24\textwidth]{\textbf{avg bpp / avg DISTS$\downarrow$}}
\makebox[0.24\textwidth]{\textbf{0.0915 / 0.0851}}
\caption{{Reconstruction results at different resolutions. On the left are the original images and on the right are the decoded results.}}
\label{flexibility}
\end{figure}

Since we use the stable diffusion for image reconstruction, some may wonder about whether our method is able to achieve image compression with different resolutions. To answer this question, we use images with different resolutions, such as 256$\times$256, 512$\times$768, and 512$\times$1538, for evaluation. As shown in Fig.~\ref{flexibility}, the proposed DiffEIC is able to reconstruct visually pleasing results under different image resolutions. In addition, we believe that our method is capable of processing ultra-high definition images (i.e., 4K and 8K) using the block-based processing strategy when the computational resources are limited.

\subsection{Limitation}
Although the proposed DiffEIC framework achieves favorable reconstructions at extremely low bitrates, it still has some limitations. 1) While text is an important component in pre-trained text-to-image diffusion models, its application has not yet been explored within our framework. The work of Text+Sketch \cite{Text+Sketch} demonstrates the powerful ability of text in extracting image semantics, encouraging us to further leverage text to enhance our method in future work. 2) Due to using a diffusion model as the decoder, the DiffEIC framework requires more computational resources and longer inference times compared to other VAE-based compression methods. Using more advanced sampling methods may be a solution to alleviate the computing burden. {3) Due to the limitations of the stable diffusion autoencoder, DiffEIC exhibits lower performance on pixel-wise distortion metrics compared to other methods. Future work will focus on improving the balance between pixel-wise accuracy and perceptual quality.}

%% file: 07_conclusion.tex
\section{Conclusion}
\label{conclusion}
In this paper, we propose a novel extreme image compression framework, named DiffEIC, which combines compressive VAEs with pre-trained text-to-image diffusion models to achieve realistic and high-fidelity reconstructions at extremely low bitrates (below 0.1 bpp). First, we introduce a VAE-based latent feature-guided compression module to adaptively select information essential for reconstruction. This module compresses images and initially decodes them into content variables. The latent feature guidance strategy effectively improves reconstruction fidelity. Second, we propose a conditional diffusion decoding module that leverages the powerful generative capability of pre-trained stable diffusion to reconstruct images with realistic details. Finally, we design a simple yet effective space alignment loss to optimize DiffEIC within a unified framework. Extensive experiments demonstrate the superiority of DiffEIC and the effectiveness of the proposed modules.

%% file: 09_biography.tex
\begin{IEEEbiography}[{\includegraphics[width=1in,height=1.25in,clip,keepaspectratio]{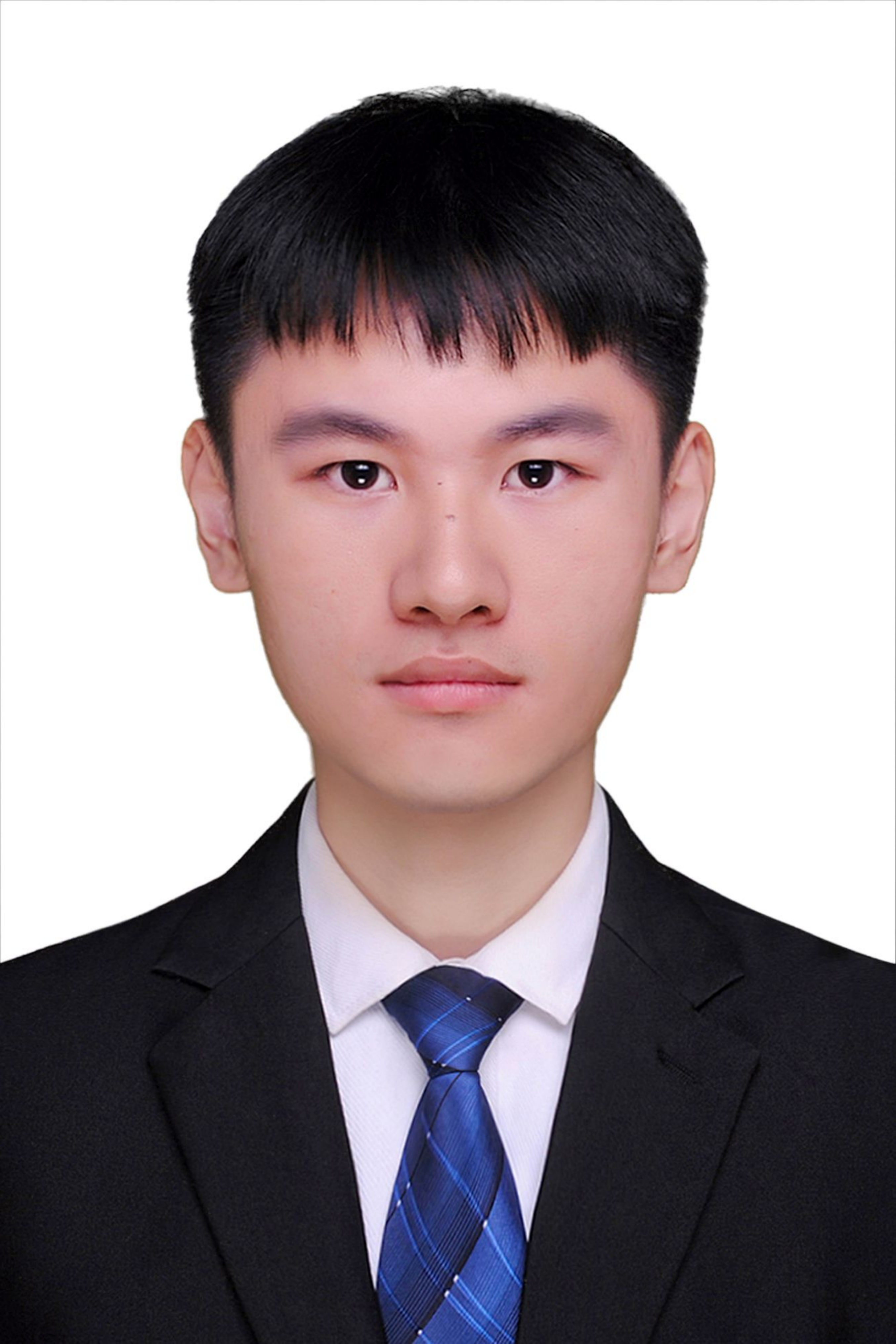}}]{Zhiyuan Li}
is currently pursuing a master's degree with the Institute of Artificial Intelligence and Robotics at Xi’an Jiaotong University. He received his bachelor's degree from Xidian University in 2022. His research interests include image compression, image rescaling, and other visual problems.
\end{IEEEbiography}

\begin{IEEEbiography}
[{\includegraphics[width=1in,height=1.25in,clip,keepaspectratio]{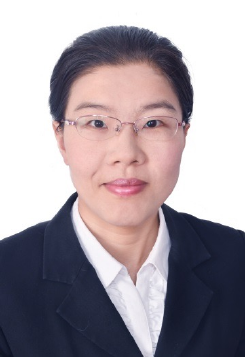}}]{Yanhui Zhou}
received the M. S. and Ph. D. degrees in electrical engineering from the Xi'an Jiaotong University, Xi'an, China, in 2005 and 2011, respectively. She is currently an associate professor with the School of Information and telecommunication Xi’an Jiaotong University. Her current research interests include image/video compression, computer vision and deep learning.
\end{IEEEbiography}

\begin{IEEEbiography}[{\includegraphics[width=1in,height=1.25in,clip,keepaspectratio]{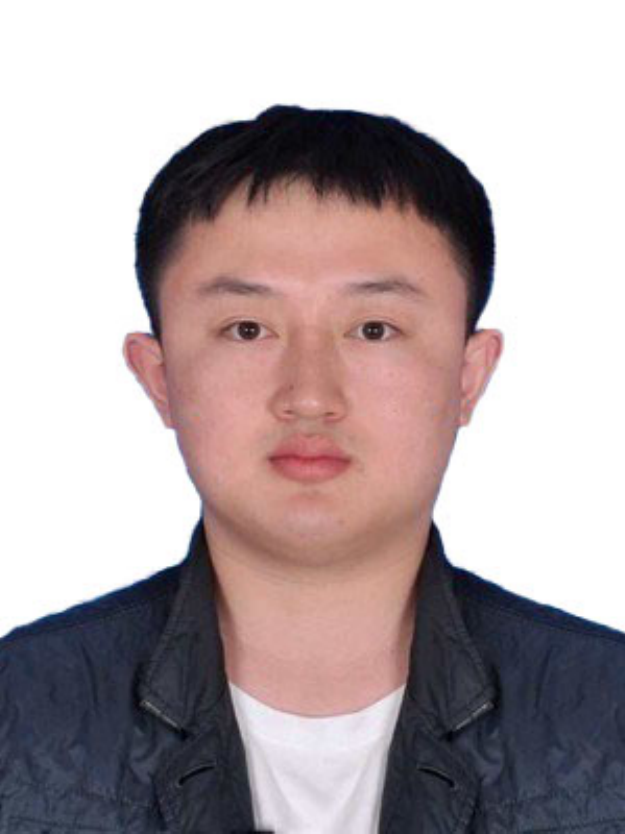}}]{Hao Wei}
is currently a Ph.D. candidate with the Institute of Artificial Intelligence and Robotics at Xi'an Jiaotong University. He received his B.Sc. and M.Sc. degrees from Yangzhou University and Nanjing University of Science and Technology in 2018 and 2021, respectively. His research interests include image deblurring, image compression, and other low-level vision problems.
\end{IEEEbiography}

\begin{IEEEbiography}[{\includegraphics[width=1in,height=1.25in,clip,keepaspectratio]{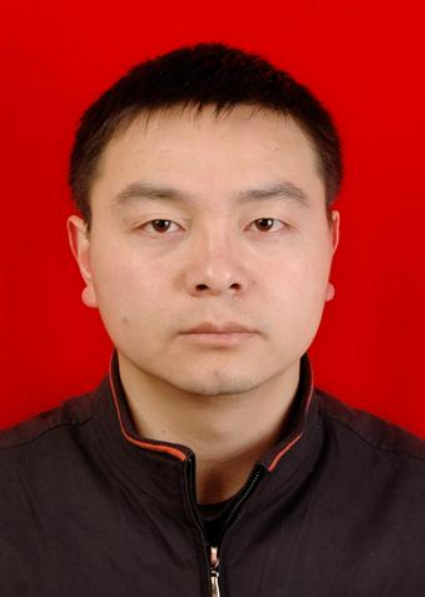}}]{Chenyang Ge}
is currently an associate professor at Xi'an Jiaotong University. He received the B.A., M.S., and Ph.D. degrees at Xi'an Jiaotong University in 1999, 2002, and 2009, respectively. His research interests include computer vision, 3D sensing, new display processing, and SoC design.
\end{IEEEbiography}

\begin{IEEEbiography}[{\includegraphics[width=1in,height=1.25in,clip,keepaspectratio]{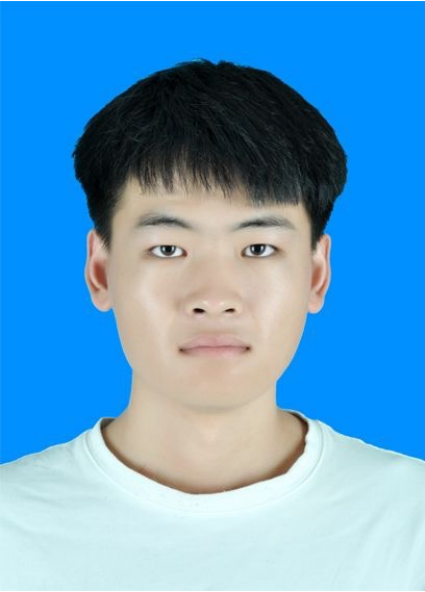}}]{Jingwen Jiang}
is currently pursuing a master's degree at Xi'an Jiaotong University. He received his bachelor's degree from Sichuan University in 2023. His research interests include image compression and video compression.
\end{IEEEbiography}